%% file: main.tex
\documentclass[sigconf, nonacm]{acmart}

\usepackage{xspace}
\usepackage{listings}
\usepackage{multirow}
\usepackage{tikz}
\usetikzlibrary{shapes,arrows}
\usepackage{pgfplots}
\usetikzlibrary{shapes.geometric, arrows}
\usepackage{subcaption}
\usepackage{algorithmicx}
\usepackage[plain]{algorithm}
\usepackage{algpseudocode}
\usepackage{array}
\usepackage{mathpartir}
\usepackage{bm}
\usepackage{amsmath}
\usepackage[inline]{enumitem}

\usepackage{silence}
\input{macros}

\begin{document}

\title{Fine-Tuning Data Structures for Analytical Query Processing}

\author{Amir Shaikhha}
\affiliation{%
  \institution{University of Edinburgh}
  \country{United Kingdom}}
\email{amir.shaikhha@ed.ac.uk}
  
\author{Marios Kelepeshis}
\affiliation{%
  \institution{University of Oxford}
  \country{United Kingdom}}
\email{marios.kelepeshis@stx.ox.ac.uk}

\author{Mahdi Ghorbani}
\affiliation{%
  \institution{University of Edinburgh}
  \country{United Kingdom}}
\email{mahdi.ghorbani@ed.ac.uk}



\begin{abstract}
We introduce a framework for automatically choosing data structures to support efficient computation of analytical workloads. Our contributions are twofold. First, we introduce a novel low-level intermediate language that can express the algorithms behind various query processing paradigms such as classical joins, groupjoin, and in-database machine learning engines. This language is designed around the notion of dictionaries, and allows for a more fine-grained choice of its low-level implementation. Second, the cost model for alternative implementations is automatically inferred by combining machine learning and program reasoning. The dictionary cost model is learned using a regression model trained over the profiling dataset of dictionary operations on a given hardware architecture. The program cost model is inferred using static program analysis.  

Our experimental results show the effectiveness of the trained cost model on micro benchmarks. Furthermore, we show that the performance of the code generated by our framework either outperforms or is on par with the state-of-the-art analytical query engines and a recent in-database machine learning framework. 
\end{abstract}

\maketitle




\input{intro}

\input{arch}

\input{langds}

\input{qopt}

\input{cost}

\input{synthesis}

\input{exp}

\input{related}

\section{Conclusion}
In this paper, we present \system, a framework that automatically synthesizes analytical engines with fine-tuned data structures.
This is facilitated by \langds{}, a low-level intermediate language based on dictionaries. 
The execution cost of \langds{} is automatically inferred by 
1) training regression models for the cost model of dictionary implementations, and 2) using cost inference rules on \langds{} statements for the cost model of the entire \langds{} program. 
Our experimental results show the effectiveness of our cost model and its performance in comparison with state-of-the-art in-memory engines for query processing and in-database machine learning.
We plan to explore including more features for the dictionary cost model. 
Another promising future direction is multi-core and parallel architectures that can impose further challenges.


\bibliographystyle{abbrv}
\bibliography{refs}

\clearpage

\appendix

\input{appendix}

\end{document}

%% file: macros.tex
\definecolor{forestgreen}{rgb}{0.13, 0.55, 0.13}

\lstdefinelanguage{llql}%
{morekeywords={
  if,then,else,let,in,%
  build,dom,true,false,key,val,%
  sum,for,iter,range,ref,%
  int,double,dense_int,bool,string,rng%
  },%
  sensitive,%
  morecomment=[l]//,%
  morecomment=[s]{/*}{*/},%
  morestring=[b]",%
  morestring=[b]',%
  showstringspaces=false,%
  morecomment=[s][\color{gray}]{@}{\ },%
    breaklines=true,%
  mathescape=true,%
showspaces=false,
showtabs=false,
showstringspaces=false,
breakatwhitespace=true,
  aboveskip=1pt,
  belowskip=1pt,
  lineskip=-0.2pt,
   numbersep=5pt,
   numberstyle=\tiny\ttfamily,
   basicstyle=\small\ttfamily,
   keywordstyle=\bfseries\color{blue!70!black},%
   columns=fullflexible,
  frame=single,
  escapeinside={(*@}{@*)}
}[keywords,comments,strings]%

\lstMakeShortInline[columns=fixed]!

\lstdefinelanguage{pseudo}%
{morekeywords={
  Dictionary%
  },%
  sensitive,%
  morecomment=[l]//,%
  morecomment=[s]{/*}{*/},%
  morestring=[b]",%
  morestring=[b]',%
  showstringspaces=false,%
  morecomment=[s][\color{gray}]{@}{\ },%
    breaklines=true,%
  mathescape=true,%
showspaces=false,
showtabs=false,
showstringspaces=false,
breakatwhitespace=true,
  aboveskip=1pt,
  belowskip=1pt,
  lineskip=-0.2pt,
   numbersep=5pt,
   numberstyle=\tiny\ttfamily,
   basicstyle=\small\ttfamily,
  keywordstyle=\small\bfseries\color{green!50!black},%
      commentstyle     = \color{forestgreen},%
  columns=fullflexible,
  frame=single,
  escapeinside={(*@}{@*)}
}[keywords,comments,strings]%

\newcommand{\langds}{LLQL\xspace}
\newcommand{\system}{DBFlex\xspace}
\newcommand{\grammarcomment}[1]{\textit{\small #1}}
\newcommand{\smartpara}[1]{\noindent \textbf{#1.}}
\newcommand{\col}[1]{\textbf{#1}}
\newcommand{\code}[1]{\texttt{#1}}

\let\bowtie\relax
\DeclareFontEncoding{LS1}{}{}
\DeclareFontSubstitution{LS1}{stix}{m}{n}
\DeclareSymbolFont{STIXsymbols}{LS1}{stixscr}{m}{n}
\DeclareMathSymbol{\bowtie}{\mathrel}{STIXsymbols}{"0E}

\def\gjoin{\setbox0=\hbox{$\bowtie$}%
  \rule[1.20ex]{.30em}{.6pt}}
\def\groupjoin{\mathbin{\bowtie\mkern-6.3mu\gjoin}}

\DeclareMathOperator*{\argmin}{argmin}
\newcommand{\alglist}[1]{\bm{#1}}

\newcommand{\stlumap}{unordered\_map}
\newcommand{\rbdict}{robinhood\_dict}
\newcommand{\tsldict}{tsl\_dict}
\newcommand{\boostumap}{boost\_unordered\_map}
\newcommand{\stlmap}{map}
\newcommand{\boostfmap}{boost\_flat\_map}
\newcommand{\tlxdict}{tlx\_dict}
\newcommand{\absldict}{absl\_dict}

\newcommand{\dsdict}[1]{\texttt{#1}}
\newcommand{\regmodel}[1]{\textsc{#1}}

\definecolor{regrmodelcolor}{RGB}{254,254,128}
\definecolor{cardmodelcolor}{RGB}{127,254,128}
\definecolor{dictimplcolor}{RGB}{127,255,255}

\newcommand{\regrmodelshape}{\begin{tikzpicture}[scale=0.1, baseline=0mm]
\draw[fill=regrmodelcolor] (5,1) node[anchor=north]{}
  -- (0,2) node[anchor=north]{}
  -- (0,0) node[anchor=south]{}
  -- cycle;
\end{tikzpicture}}

\newcommand{\cardmodelshape}{\begin{tikzpicture}[scale=0.1, baseline=0mm]
\draw[fill=cardmodelcolor] (5,0) node[anchor=north]{}
  -- (5,2) node[anchor=north]{}
  -- (0,2) node[anchor=north]{}
  -- (0,0) node[anchor=south]{}
  -- cycle;
\end{tikzpicture}}

\newcommand{\dictimplshape}{\begin{tikzpicture}[scale=0.1, baseline=0mm]
\draw[fill=dictimplcolor] (5,0) node[anchor=north]{}
  -- (5,2) node[anchor=north]{}
  -- (1,2) node[anchor=north]{}
  -- (0,1) node[anchor=south]{}
  -- (1,0) node[anchor=south]{}
  -- cycle;
\end{tikzpicture}}

%% file: intro.tex
\section{Introduction}
Query processing engines have undergone a massive progress over the previous decade.
Traditionally, the volcano iterator model~\cite{Volcano} was considered the de-facto standard for building pipelined query engines. 
This model streams the data along the query operators, and works well for out-of-core scenarios.
For in-memory databases, the runtime overhead caused by this iterator model is mitigated by a mixture of techniques such as query compilation~\cite{krikellas, Neumann11, dbtoaster, legobase_tods, DBLP:journals/debu/ViglasBN14, crotty2015tupleware, Nagel:2014:CGE:2732977.2732984, karpathiotakis2015just, spark-sql} and 
vectorization~\cite{monetdb-handwritten,Polychroniou:2015:RSV:2723372.2747645,Zhou:2002:IDO:564691.564709}.

To accommodate specialized query operators in the design of modern query processing engines for in-memory databases, the following considerations are common.

\smartpara{Hash-Based and Sort-Based Query Operators}
The efficient evaluation of query operators can benefit from
hash-based and sort-based data-structures~\cite{10.1145/2882903.2912569,cowbook,Graefe:1993:QET:152610.152611}.
The trade-offs between hashing and sorting has been investigated in depth in the literature~\cite{kim2009sort,albutiu2012massively,balkesen2013multi,mirzadeh2015sort,muller2015cache}.
Most database systems tend to implement various types of physical query operators using these two approaches (e.g., sort-merge-join and hash-join), and delegate to the optimizer the task of picking the best choice based on the workload features.

\smartpara{Specialized Compound Query Operators}
Such operators may be beneficial for OLAP workloads and are implemented in state-of-the-art in-memory database systems~\cite{neumann2017complete,Menon:2017:ROF:3151113.3151114}.
For example, the groupjoin operator~\cite{groupjoin} merges aggregate and hash-join operators.

The above considerations typically lead to two main challenges for the query optimizer.
First, for every specialized query operator, the database developer needs to extend the set of supported query operators.
Second, one has to provide a cost model for the new query operator. This can be daunting especially if the difference is only in the low-level implementation details for the data structures.

This paper addresses these issues by proposing \system{}, a query processing engine with two main design decisions.
First, \system{} uses a \textbf{dictionary-based intermediate language}.
This language is expressive enough to capture query processing algorithms for, e.g., classical query operators, compound operators, and in-database machine learning engines, and allows a cost-based choice for its dictionary implementation. 

Second, \system{} \textbf{automatically infers the cost models} for alternative implementations of a query and may uncover the right trade-off between hashing and sorting for operators based on the given workload and its underlying hardware architecture. 
Our design thus frees the database developers from the difficult and error-prone task 
of defining a cost model for different query operators.

\noindent \textbf{Motivating Example.} To better understand the differences from previous approaches, consider the following simplified TPCH query Q3, where we removed one join and simplified the group-by clause:

\begin{lstlisting}[language=SQL,frame=none,keywordstyle=\small\ttfamily\bfseries]
  select L.K, sum(L.P * L.D) 
  from L join O on L.K = O.K 
  where O.T < %DATE%
  group by L.K
\end{lstlisting}

\noindent
The join on !K! is key/foreign-key. 
Following TPCH specifications, relation !O! cannot be indexed by !T!. As !K! is a part of compound key for relation !L!, this relation can be ordered on it.
Most traditional query engines process the query using two hash-tables.
The first hash table is built and probed to construct the intermediate join result.
The second hash table is used for the group-by aggregate result.
The pseudocode for the corresponding data-centric compiled engine~\cite{Neumann11} is as follows:

\begin{lstlisting}[language=pseudo,frame=none]
  init (*@$\Gamma_{HT}$@*), (*@$\bowtie_{HT}$@*) as HashTable
  for each tuple l in L
    (*@$\bowtie_{HT}$@*).insert(l.K, l)
  for each tuple o in O
    if o.T < %DATE%
      if((*@$\bowtie_{HT}$@*).contains(o.K))
        for each tuple l in (*@$\bowtie_{HT}$@*)[o.K]
          (*@$\Gamma_{HT}$@*)[l.K] += l.P * l.D
  return (*@$\Gamma_{HT}$@*)
\end{lstlisting}

\noindent
This evaluation strategy can be improved as follows:

\noindent
\textit{1) Compound Groupjoin.} In this query, the join key and the group-by attribute are the same.
Thus, the intermediate hash tables $\Gamma_{HT}$ and $\bowtie_{HT}$ can be merged. The resulting
compound operator is referred to as groupjoin~\cite{groupjoin}.
As a result, the previous query can be rewritten as follows:

\begin{lstlisting}[language=pseudo,frame=none]
  init (*@$\groupjoin_{HT}$@*) as HashTable
  for each tuple o in O
    if o.T < %DATE%
      (*@$\groupjoin_{HT}$@*)[o.K] = 0
  for each tuple l in L
    if((*@$\groupjoin_{HT}$@*).contains(l.K))
      (*@$\groupjoin_{HT}$@*)[l.K] += l.P * l.D
  return (*@$\groupjoin_{HT}$@*)
\end{lstlisting}

\begin{figure}[t]
\includegraphics[width=\columnwidth]{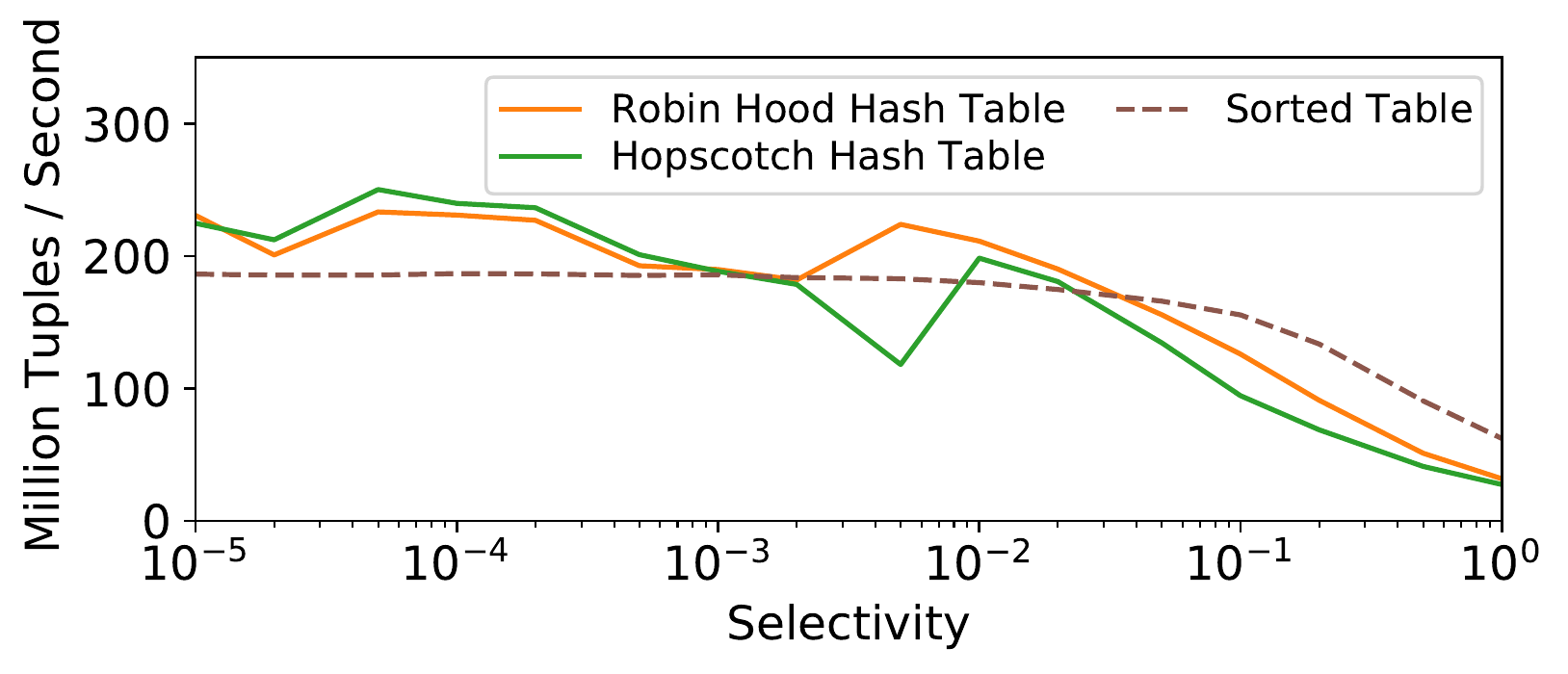}
\caption{Performance of the query in the running example with different dictionaries as function of the selectivity of the predicate on \texttt{O.T}. 
Relation \texttt{L} is already ordered on \texttt{K}.}
\label{fig:gj}
\end{figure}

\noindent
\textit{2) Specialized Hash-Tables.} The intermediate hash table for join, group-by 
aggregate, and groupjoin can be implemented in various ways. As an example, to resolve
hash collisions there are different approaches such as Hopscotch~\cite{herlihy2008hopscotch} and Robin hood~\cite{celis1985robin} hashing.
Each implementation can be beneficial for different selectivities (cf. Figure~\ref{fig:gj}).

\noindent
\textit{3) Sort-Based Dictionaries.} Apart from using hash tables, one can use sort-based
dictionaries in order to maintain the intermediate joins and aggregates. Examples are 
tree-base dictionaries (e.g., $B^+$-trees, Red-Black trees, etc.) and sorted arrays.
These dictionaries can be especially useful when one of the input relations is already 
ordered based on the join/group-by key, which is the case for relation !L! in our running example.
Similarly, one can also have a sort-based variant of groupjoin operator, which is used
in engines such as LMFAO~\cite{Schleich:2019:LAE:3299869.3324961} and FDB~\cite{Olteanu:2016:FD:3003665.3003667}.

In order to support these specialized operators we need to overcome the following challenges.
First, the database developer needs to extend the set of query operators with hash-based and
sort-based groupjoin, each variant with possibly different specialized implementations.
Second, designing the cost model for each variant is very tedious and is not easily portable
to different hardware architectures.

\system solves these issues by introducing an intermediate language around dictionaries. 
First, the program in this language does not specify the data structure for dictionary $\groupjoin_{Dict}$. Nevertheless, it encodes the join order as well as the basic algorithm behind 
groupjoin. Hence, \textit{there is no need to extend the set of query operators}. 
More specifically, the query is expressed as follows:

\begin{lstlisting}[language=pseudo,frame=none]
  init (*@$\groupjoin_{Dict}$@*) as Dictionary
  for each tuple o in O
    if o.T < %DATE%
      (*@$\groupjoin_{Dict}$@*)[o.K] = 0
  for each tuple l in L
    if((*@$\groupjoin_{Dict}$@*).contains(l.K))
      (*@$\groupjoin_{Dict}$@*)[l.K] += l.P * l.D
  return (*@$\groupjoin_{Dict}$@*)
\end{lstlisting}

\noindent 
Then, by using program synthesis, the implementation of the underlying dictionary becomes explicit.
To estimate the run time of the program for each dictionary implementation, \system uses \textit{machine learning} and \textit{program reasoning}. First, to model the cost of dictionaries, a regression model is trained 
whose input features are the size of dictionary, the orderedness of input data, and the number of accessed tuples.
This way, there is no need to use any hardware parameters as features, as \system 
profiles the dictionary operations on each machine.
Second, \system statically analyzes the statements to estimate the execution time of the whole program.

Figure~\ref{fig:gj} shows that the best dictionary implementation depends on the selectivity: 1) this is hopscotch hashing for selectivities under $0.1\%$, 
2) it is robin hood hashing for selectivities between $0.1\%$ and $5\%$, 
and 3) becomes sorted table for selectivities larger than $5\%$, as its amortized lookup cost starts paying off.
\system uses dictionary size and number of access tuples, which define the selectivity.

\input{figures/archfig}

The contributions of this paper are as follows:

\begin{itemize}[leftmargin=*]
\setlength\itemsep{0em}
    \item We propose a new architecture for building database systems using our proposed intermediate language (Section~\ref{sec:arch}). The high-level view of the architecture of \system is shown in Figure~\ref{fig-arch}. \system allows for defining specialized dictionary implementations, the cost models of which can be automatically learned. 
    \item We introduce a novel low-level intermediate language, called \langds{}. Our intermediate language is designed around the notion of nested dictionaries~\cite{shaikhha2021functional} that generalize flat relations, nested relations~\cite{roth1988extended}, tree-based indices, and trie-based representations~\cite{Olteanu:2016:FD:3003665.3003667} 
    (Section~\ref{sec:langds}).
     \langds{} can express various basic query operators (selection, projection, aggregation, and nested loop join). Furthermore, it can express physical operators (hash-based and sort-based groupby, hash join, sort-merge join, and index-nested loop join), compound operators (hash-based and sort-based group\-join), and efficient in-DB machine learning engines.
    \item \langds{} is designed with two goals in mind. First, it has to be \textit{low-level enough} to capture the underlying hardware architecture behavior (e.g., execution time of dictionary operations). 
    Second, it should be \textit{high-level enough} to allow for statically reasoning about the programs (e.g., run-time execution cost of programs).
    We show how the low-level nature of \langds{} allows us to use regression models learned over dictionary-related features such as dictionary size, number of operator invocations, and orderedness of data to capture the dictionary cost model (Section~\ref{sec:cost:dict}).
    Furthermore, thanks to the domain-specific nature of \langds{}, we show how use program reasoning to infer the cost of \langds{} expressions by using the dictionary cost model and a cardinality model (Section~\ref{sec:costinfer}). 
    \item The derived cost model (Section~\ref{sec:cost}) allows \system{} to automatically synthesize the \langds{} program with the best cost.
    We present a greedy algorithm for choosing the dictionary implementations that lead to the \langds{} with the lowest cost estimate (Section~\ref{sec:ds_synthesis}).
    \item Finally, we show experimentally the effectiveness of the learned dictionary cost model and the inferred \langds{} cost model. 
    Also, we show the advantage of using several dictionary implementations for a query 
    over using a single dictionary. Overall, our engine outperforms the state-of-the-art engines Typer and Tectorwise by 1.5x and respectively 2x in average, while also recovering the runtime performance of the LMFAO in-database machine learning engine, which is tuned for specific workloads (Section~\ref{sec:exp}).
    
\end{itemize}

%% file: figures/archfig.tex
\begin{figure}[t!]
\includegraphics[width=\columnwidth]{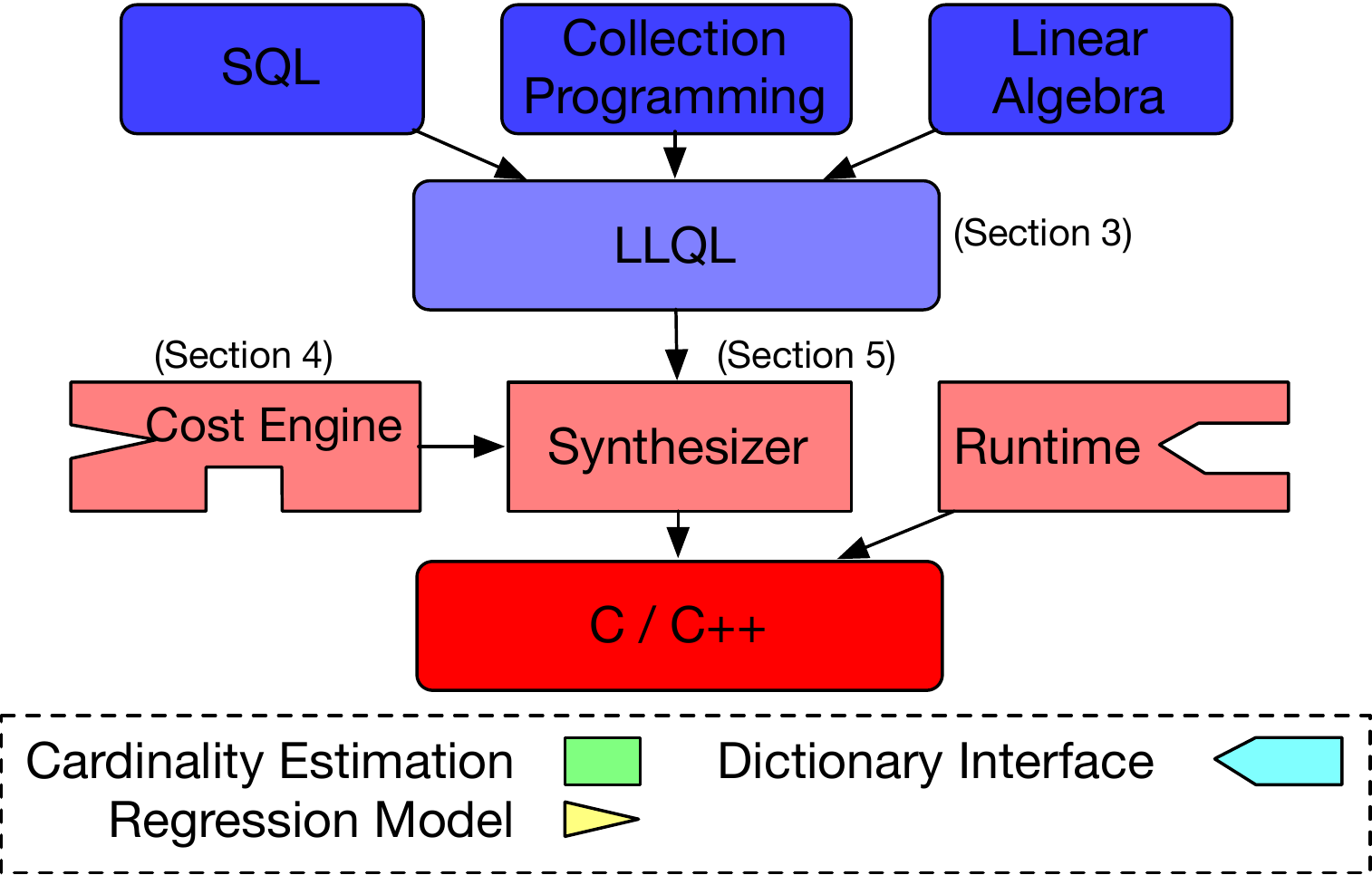}
\caption{High-level architecture of \system.}
\label{fig-arch}
\end{figure}

%% file: arch.tex
\section{Architecture and System Design}
\label{sec:arch}

In this section, we present the architecture of \system. First, we describe the high-level
architecture of our system (Section~\ref{sec:arch_hl}). 
Then, we describe the workflow of our cost-based program synthesis (Section~\ref{sec:arch_wf}).
Finally, we show how database developers can extend \system with alternative dictionary implementations (Section~\ref{sec:arch_ext}).

\subsection{Overall System Architecture}
\label{sec:arch_hl}

The architecture of \system is presented in Figure~\ref{fig-arch}.
The input program can be in a variety of languages including SQL, linear algebra, and functional collection programming languages.
This means that \system can serve as the backend engine of existing DBMSes by getting their produced query plan and generating the optimized C++ code for it.
Futhermore, \system can not only run a wide range of analytical workloads, but also hybrid workloads such as in-database machine learning.

As opposed to optimizing queries at the level of physical query plan, \system goes \textit{deeper}~\cite{dittrich2019case}.
Given a dataflow of query operators (i.e., the join order is specified using state-of-the-art techniques~\cite{yan2019generating,trummer2017solving}), \system synthesizes a \langds{} program with the dictionary implementations that lead to lowest cost estimates. 

Similar to query optimization, there are two main components for the synthesis: 
\begin{itemize}[leftmargin=*]
\item \textbf{Cost Model:} \system defines cost model at the level of dictionary operations and \langds{} program. The dictionary cost model is learned using a regression model over the profiling data (Section~\ref{sec:cost:dict}). The \langds{} cost model is statically inferred using inference rules over \langds{} constructs (Section~\ref{sec:costinfer}).
\item \textbf{Search:} \langds{} synthesis can be implemented using the same search techniques 
employed in query optimizers such as dynamic-programming, randomized algorithms, or other optimization techniques such as integer linear programming~\cite{yan2019generating,trummer2017solving}).
Section~\ref{sec:ds_synthesis} gives a greedy algorithm for this component.
\end{itemize}

\subsection{Workflow}
\label{sec:arch_wf}

Figure~\ref{fig-workflow} shows the workflow of \system in three stages.\footnote{In practice, the query optimization and query execution are the same stages. We make this distinction for the sake of presentation.} 

\noindent \textbf{1) Installation Stage:} At this stage, the database system is being deployed to a particular machine. 
By generating a synthetic profiling data and running the operations of various dictionary implementations we generate a training set. 
Then, we can train a regression model to capture the dictionary cost model.

\noindent \textbf{2) Query Optimization:} The input query is provided at this stage,
and is translated to a logical execution plan with the choice for join orders (e.g., by Postgres or Apache Calcite).
The logical plan is translated into \langds{} without the implementation choice for dictionaries.  
The program synthesis generates the search space of different \langds{} programs with different dictionary implementations. By using the trained regression models and inferring the execution cost of alternative \langds{} programs (Section~\ref{sec:costinfer}), the \langds{} with the best dictionary implementations is chosen.
The best \langds{} program is translated to low-level specialized engine code in C++.

\noindent \textbf{3) Query Execution:} Finally, the input data is passed to the generated specialized engine and output result is produced.

\input{figures/workflow}

\subsection{Extensions}
\label{sec:arch_ext}

DB developers can extend \system in three dimensions:

\begin{itemize}[leftmargin=*]
\item \textbf{Dictionary Implementation:} The runtime of \system includes a dictionary interface serving as an extension point (cf. Figure~\ref{fig-workflow}).
\system exposes the API shown in Figure~\ref{fig:dictapi}, inspired by the API of standard library of C++ for dictionaries. 
After providing an appropriate hash-based or sort-based dictionary implementation, the developer needs to register the dictionary to be used during the installation.
\item \textbf{Regression for Dictionary Cost Model:} \system uses out-of-the-box regression models provided by machine learning frameworks (e.g., sklearn and TensorFlow). 
The developers can provide additional regression models.
\item \textbf{Cardinality Model:} Cardinality and selectivity estimates are essential components for cost-based optimizers. \system relies on state-of-the-art cardinality estimation models~\cite{getoor2001selectivity,moerkotte2009preventing,yang13deep}, as studying the impact of different cardinality estimation models is beyond the scope of this paper. However, the developer can use alternative cardinality models. 
\end{itemize}

\input{figures/dictapi}

%% file: figures/workflow.tex
\begin{figure}[t!]
\includegraphics[width=\columnwidth]{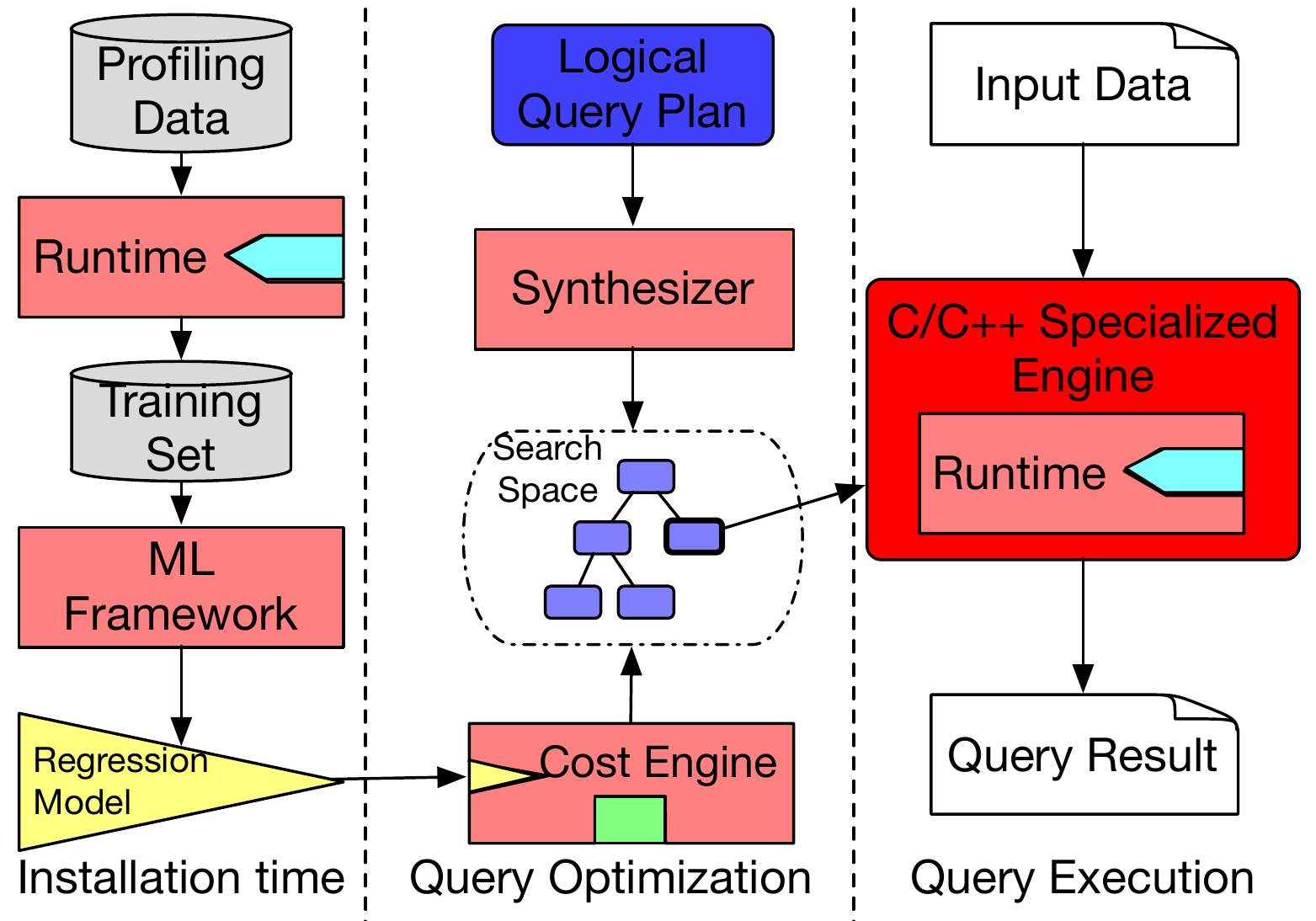}
\caption{The workflow of query processing in \system.}
\label{fig-workflow}
\end{figure}

%% file: figures/dictapi.tex
\begin{figure}[t]
\hfill\includegraphics[width=0.12\columnwidth]{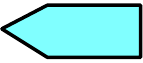}\\
\vspace{-0.6cm}
\begin{lstlisting}[language=C++]
template<class Key, class Value>
class dict_impl {
  class iterator; /* Iterator interface */
  iterator begin();
  iterator end();
  iterator find(const Key& k);
  iterator emplace(const Key& k, const Value& v);
  /* Only for sorted dictionaries */
  iterator find_hint(iterator position, const Key& k);
  iterator emplace_hint(iterator position, 
                          const Key& k, const Value& v);
};
\end{lstlisting}
\caption{The dictionary interface exposed by the runtime of \system.}
\label{fig:dictapi}
\end{figure}

%% file: langds.tex
\section{\langds}
\label{sec:langds}
\langds{} is a domain-specific language inspired by bag-based and dictionary-based query languages (e.g., AGCA~\cite{dbtoaster} and FAQ~\cite{abo2016faq}).
There are two major design decision behind this language.
First, \langds{} is \textit{not a purely functional language}; 
in order to have full control on performance, the dictionary data-structure is not immutable. This means that one can destructively update the value associated with a key without the need to recreate another dictionary.
Second, \langds{} can be data-structure-aware and encode the implementation choice for the underlying dictionaries.
More specifically, the dictionary accesses in this language are either based on hash tables
or sorted data-structures.
\langds{} can express physical operators (hash-based and sort-based groupby, hash join, sort-merge join, and index-nested loop join), compound operators (hash-based and sort-based group\-join), and efficient in-DB ML engines.

\subsection{Syntax}
Figure~\ref{fig:langds} shows the grammar of \langds for both expressions (!e!) and types (!T!).
The core data type supported by \langds is a dictionary, represented as !{{ T -> T }}!.
\langds represents input relations as dictionaries from tuples to their multiplicity, because
of their bag semantics in database systems, as opposed to their set semantics in relational algebra. 

The expression !for(r <- R) e! specifies iteration over the elements of dictionary !R!,
and performing !e! at each iteration.
Records can be constructed using !{ a_1 = e_1, ..., a_n = e_n }! and
the field !a_i! of record !rec! can be accessed using !rec.a_i!.

\input{figures/grammards}

\input{figures/qopts}

\subsection{Dictionaries}
The !{{ k -> v }}! constructs a singleton dictionary that has !k! and !v! as its key and value, respectively.
The !dict(k)! operator performs a lookup for key !k! in the dictionary !dict!.
The elements of a dictionary are key-value pairs, which can be seen as records with 
field names !key! and !val!.
Thus, in the body of the loop !for(r <- R)!, one has to use !r.key! and !r.val! to access the key and value of !r!, respectively.
The addition for dictionaries is defined in terms of elementwise addition; the values with the same key are added.

The choice of data structure is specified using !@ds!, which is used for constructing a dictionary: !@ds {{ e -> e }}!.
We can use every specialized dictionary data structure instead of !@ds!. However,
for the sake of brevity, we only use these two annotations: 1) !@ht! for hash-based dictionaries, and 2) !@st! for sort-based dictionaries. Next, we present these two types of dictionaries. 

\subsubsection{Hash-based Dictionaries}
One of the most obvious ways of implementing a dictionary is by using 
a hash-table data-structure.
The keys are first mapped through a hash function to a particular bucket
and the associated values can be accessed, inserted, or removed with a constant-time complexity.

One of the key challenges for hash-based dictionaries is handling collisions.
Database systems have developed different approaches such as robin hood hashing and chained hashing~\cite{richter2015seven}.
\langds{} can make such choices explicit by using an appropriate hash table implementation.

The operators provided for a hash-based dictionary are: 1) iterating over 
a dictionary !dict! is specified using !for(x <- dict)!, 2) inserting or 
updating the value !v! associated with the key !k! in a dictionary !dict! is 
specified using !dict += k -> v!, and 3) looking up the value associated 
with key !k! in a dictionary !dict! is specified using !dict(k)!.\footnote{\langds{} can be extended to support deletions as well. This makes it
appropriate for transactional workloads, which we leave for the future.
}

\subsubsection{Sort-Based Dictionaries}
In databases, there has been always a dual approach to hashing, which is based on sorted data structures~\cite{cowbook}.
Such data structures can be either implemented using sorted collections, or
through tree-based data structures such as B-Trees, B$^+$-Trees, Red-Black Trees, etc.

Dictionaries that are implemented using such sort-based data structures can achieve logarithmic time for access, insertion, and updates.
Similar to hash-based dictionaries, the sort-based dictionaries have the following three operations: 1) iteration using !for(x <- dict)!, 
2) insertion or update using !dict += k -> v!, and 3) look up using !dict(k)!.
Each of these operators can leverage the fact that the dictionary is already ordered. 
For example, the lookup can use various specialized algorithms in order to further benefit from the underlying architecture~\cite{inmem-sorted-search}.

Furthermore, if the accessed keys are ordered, one can 
lower the logarithmic run times to constant time.
In order to access the values associated with a key, if one knows that the accessed keys are ordered, 
there is no need to lookup the whole range at each iteration.
Instead, one can only consider the keys not already covered by the previous iterations.
The same technique can be applied to insertion.

In order to support these optimized operators, \langds{} provides facilities in order to 
retrieve the iterator of a collection.
There are three constructs related to this feature: 1) the iterator of the dictionary !dict! is retrieved using !dict.iter!,
2) the iterator !it! is used as a hint for lookup of the key !k! in the dictionary !dict! using !dict<it>(k)!, and 3) the iterator !it! is used as a 
hint for updating the value associated with the key !k! in the dictionary !dict! by !v! using !dict<it> += k -> v!. 
Note that both lookup and insert operators update the passed iterator.

\subsubsection{Mapping to Runtime}
The iteration over a dictionary corresponds to the \code{begin()} and \code{end()} functions of the Runtime API in Figure~\ref{fig:dictapi}. The lookup corresponds to the \code{find()} function. The update construct is implemented by first invoking \code{find()}. 
In the case that a match is found, the corresponding value is incremented by the input value. Otherwise, the corresponding key-value pair is inserted by using the \code{emplace()} function.
Finally, the hinted versions correspond similarly to \code{find\_hint()} and \code{emplace\_hint()}.

Next, we show how both these two data structures can be used for implementing various physical query operators.

%% file: figures/grammards.tex
\begin{figure}[t]
\begin{tabular}{|l c l|p{0.372\columnwidth}|}
\hline
\multicolumn{3}{|c|}{\textbf{Grammar}} & \multicolumn{1}{c|}{\textbf{Description}}\\\hline
!e! & \mbox{::=} & !e ; e! & \grammarcomment{Sequence of Statements}\\
& $\mid$ & !()! & \grammarcomment{No-op}\\
& $\mid$ & !let x = e in e! & \grammarcomment{Variable Binding}\\
& $\mid$ & !if(e) then e else e! & \grammarcomment{Conditional}\\
& $\mid$ & !{ a = e }! & \grammarcomment{Record Construction}\\
& $\mid$ & !e.a! & \grammarcomment{Field Access}\\
& $\mid$ & !e b_op e! $\mid$ !u_op e! & \grammarcomment{Binary \& Unary Operations}\\
& $\mid$ & !n! $\mid$ !r! $\mid$ !false! $\mid$ !true! & \grammarcomment{Numeric \& Boolean Values}\\
& $\mid$ & !"some_text"! & \grammarcomment{String Literal}\\
& $\mid$ & !ref(T)! & \grammarcomment{Mutable Reference Init}\\
& $\mid$ & !e += e! & \grammarcomment{Mutable Reference Update}\\
& $\mid$ & !@ds {{e -> e}}! & \grammarcomment{Dictionary Construction}\\
& $\mid$ & !for(x <- e) e! & \grammarcomment{Dictionary Iteration} \\
& $\mid$ & !e(e) += e! & \grammarcomment{Dictionary Update}\\
& $\mid$ & !e(e)! & \grammarcomment{Dictionary Lookup}\\
& $\mid$ & !e.iter! & \grammarcomment{Dictionary Iterator}\\
& $\mid$ & !e<e>(e) += e! & \grammarcomment{Dictionary Hinted Update}\\
& $\mid$ & !e<e>(e)! & \grammarcomment{Dictionary Hinted Lookup}\\
\hline
!T! & \mbox{::=} & !@ds {{ T -> T }}! & \grammarcomment{Dictionary Type} \\
& $\mid$ & !int! $\mid$ !double! $\mid$ !bool! & \grammarcomment{Numeric \& Boolean Types}\\
& $\mid$ & !string! & \grammarcomment{String Type}\\
\hline
!@ds! & \mbox{::=} & !@ht! & \grammarcomment{Hash-Table Dictionary}\\
& $\mid$ & !@st! & \grammarcomment{Sort-Based Dictionary}\\
\hline
\end{tabular}
\vspace{0.1cm}
\caption{Grammar of \langds{}.}
\vspace{-0.4cm}
\label{fig:langds}
\end{figure}

%% file: figures/qopts.tex
\begin{figure*}[t]
\newcommand{\afspace}{\vspace{-0.1cm}}
\newcommand{\bfspace}{\vspace{0.2cm}}
\begin{minipage}{0.48\textwidth}
\begin{lstlisting}
let Sh = @ht {{ }} in
for(s <- S) {
  Sh( part(s.key) ) += @ht {{ s.key -> s.val }}
} ;
let RS = @ht {{ }} in

for(r <- R) {
  let rkey = part(r.key) in
  for(s <- Sh(rkey)) {
    RS( concat(r.key, s.key) ) += r.val * s.val
} }
\end{lstlisting}
\afspace
\subcaption{Hash join of \texttt{R} and \texttt{S}, join key given by \texttt{part}.}
\label{fig:qopt_hj}
\bfspace
\end{minipage}~\hspace{0.3cm}~%
\begin{minipage}{0.48\textwidth}
\begin{lstlisting}
let Ss = @st {{ }} in
for(s <- S) {
  Ss( part(s.key) ) += @st {{ s.key -> s.val }}
} ;
let RS = @st {{ }} in
let it = Ss.iter in
for(r <- R) {
  let rkey = part(r.key) in
  for(s <- Ss<it>(rkey)) {
    RS( concat(r.key, s.key) ) += r.val * s.val
} }
\end{lstlisting}
\afspace
\subcaption{Sort-merge join of \texttt{R} and \texttt{S}, join key given by \texttt{part}.}
\label{fig:qopt_sj}
\bfspace
\end{minipage}

\begin{minipage}{0.48\textwidth}
\begin{lstlisting}
let Ragg = @ht {{ }} in

for(r <- R) {
  Ragg( grp(r.key) ) += agg(r.key) * r.val
}
\end{lstlisting}
\afspace
\subcaption{Hash-based aggregation \texttt{agg} grouped by \texttt{grp}.}
\label{fig:qopt_ha}
\bfspace
\end{minipage}~\hspace{0.3cm}~%
\begin{minipage}{0.48\textwidth}
\begin{lstlisting}
let Ragg = @st {{ }} in
let it = Ragg.iter in
for(r <- R) {
  Ragg<it>( grp(r.key) ) += agg(r.key) * r.val
}
\end{lstlisting}
\afspace
\subcaption{Sort-based aggregation \texttt{agg} grouped by \texttt{grp}.}
\label{fig:qopt_sa}
\bfspace
\end{minipage}

\begin{minipage}{0.48\textwidth}
\begin{lstlisting}
let Aggh = @ht {{}} in
let Sh = @ht {{}} in
for(s <- S) {
  Sh( s.key.A ) += g(s)
} ;

for(r <- R) {
  for(gs <- Sh(r.key.A)) {
    Aggh( r.key.A ) += f(r) * gs.val
} }
\end{lstlisting}
\afspace
\subcaption{Hash-based groupjoin on \texttt{A} with partial aggregates \texttt{f} and \texttt{g}.}
\label{fig:qopt_hgj}
\end{minipage}~\hspace{0.3cm}~%
\begin{minipage}{0.48\textwidth}
\begin{lstlisting}
let Aggs = @st {{}} in
let Ss = @st {{}} in
for(s <- S) {
  Ss( s.key.A ) += g(s)
} ;
let it1 = Ss.iter
let it2 = Aggs.iter
for(r <- R) {
  for(gs <- Ss<it1>(r.key.A)) {
    Aggs<it2>( r.key.A ) += f(r) * gs.val
} }
\end{lstlisting}
\afspace
\subcaption{Sort-based groupjoin on \texttt{A} with partial aggregates \texttt{f} and \texttt{g}.}
\label{fig:qopt_sgj}
\end{minipage}
\vspace{0.1cm}
\caption{Different query operators as \langds{} expressions.}
\label{fig_qopts}
\vspace{-0.5cm}
\end{figure*}

%% file: qopt.tex
\subsection{Basic Query Operators}
\subsubsection{Selection}
Consider a relation !R!, for which we are interested in selecting the elements that satisfy a predicate !p!.
For each element !r! of this relation, if the predicate is satisfied, we increment the multiplicity of the associated value with the key !r.key! (which specifies the row of the relation) by !r.val! (which specifies the multiplicity of that row).
Otherwise, we do nothing.

\begin{lstlisting}
let sel = {{}} in
for(r <- R) {
  if(p(r.key)) then  sel(r.key) += r.val
  else               ()
}
\end{lstlisting}

\subsubsection{Projection} Similar to selection, we iterate over the elements of the relation !R!. 
This time, we update the element of the dictionary specified the application of the projection function !f! on each row of relation (!f(r.key)!) as its key and unchanged value (!r.val!).

\begin{lstlisting}
let proj = {{}} in
for(r <- R) {
  proj(f(r.key)) += r.val
}
\end{lstlisting}

\subsubsection{Nested-Loop Join}
\label{sec:nlj}
For this operation, we have to use nested loops iterating over the elements !r! and !s! of relations !R! and !S!, respectively.
For each combination of tuples, we check if !joinCond! is satisfied.
If this is the case, we update the dictionary with an element that has the combination of the tuples of these two relations using function !concat! as its key, and !r.val * s.val! as its value.

\begin{lstlisting}
let join = {{}} in
for(r <- R) {
  for(s <- S) {
    if(joinCond(r.key, s.key)) then
      join( concat(r.key, s.key) ) += r.val * s.val
    else
      ()
} }
\end{lstlisting}

\noindent This expression is inefficient because one has to consider all combinations of !r! and !s!.
This situation can be improved by leveraging data locality as will be shown in Section~\ref{sec:qopt:pjoin}.

\subsubsection{Scalar Aggregation} 
These operators can be implemented by
iterating over the elements of the relation and computing the appropriate aggregate function !aggFun!. 
For example, in the case of summing the attributes !A!, !aggFun(r.key)! is replaced by !r.key.A!, and in the case of counting, is replaced by !1!.
As there could be duplicates of an element in the input relation (the multiplicity of which is shown by !r.val!), the aggregate result for each element needs to be multiplied by !r.val!.

\begin{lstlisting}
let agg = ref(double) in
for(r <- R) {
  agg += aggFun(r.key) * r.val
}
\end{lstlisting}

\subsubsection{Group-by Aggregate} The key difference between this operator and its non-grouped variant is that 
at each iteration, a group-by aggregate returns a single dictionary with the key specified by the grouping function !grp!, and the value specified using the aggregate function !aggFun!:

\begin{lstlisting}
let Agg = {{}} in
for(r <- R) {
  Agg( grp(r.key) ) -> aggFun(r.key) * r.val }}
}
\end{lstlisting}

\noindent Both scalar and group-by aggregate operators can be generalized to compute other forms of aggregates such as minimum and maximum by supplying appropriate addition and multiplication operators. 
For example, in the case of maximum, the maximum and numerical addition need to be supplied as the addition and multiplication operators, respectively~\cite{mohri2002semiring}.
To compute aggregates such as average, one has to compute both summation and count.

\subsection{Partitioned Joins}
\label{sec:qopt:pjoin}

In the case of equality joins, one can partition the elements of the relations based on their join key,
and then only combining the elements of the two relations that fall 
into the same partition.

\subsubsection{Hash Join}
Using a hash-table data-structure for a partitioned join results in a 
hash-join operator.
For example, the partitioned join between relations !R! and !S! is 
lowered to hash join in Figure~\ref{fig:qopt_hj}.

\subsubsection{Sort-Merge Join}
In a partitioned join, if one uses a sort-based dictionary rather than a hash table.
In this case, if one of the relations is ordered on the join key, one can avoid searching 
the entire range from scratch for the next matching partition, by using the hinted lookup operator.

As shown in Figure~\ref{fig:qopt_sj}, the iterator !it! is first set to the beginning of the dictionary !Ss!. 
At the first iteration over relation !R!, the hinted lookup !Ss<it>(rkey)! searches the entire range to find the next matching element. 
After returning the matching element (or an empty dictionary if it does not find any match), the iterator !it! is updated to the least upper bound position.
This limits the next lookups to the ranges that have been explored before.
Thus, these hinted lookups have an amortized constant time.

This algorithm has the same behavior as a sort-merge join operator. 
More generally, in the case where the relation !R! is not ordered based on the join key, one needs to use the partitioned join operator that
partitions (by sorting) relation !R! based on the join key.

Furthermore, when relation !S! is already sorted based on the join key, one can use a hinted insert operator.
This brings the computational complexity of the build phase from $O(n.log(n))$ to $O(n)$. 
This algorithm is essentially the same as merge join, when the input relations are already sorted:

\begin{lstlisting}
let Ss = @st {{ }} in
let it1 = Ss.iter in
for(s <- S) {
  Ss<it1>( part(s.key) ) += (@ht {{ s.key -> s.val }})
} ;
... // same as Figure(*@ \ref{fig:qopt_sj} @*)
\end{lstlisting} 

\subsubsection{Tree-Based Join}
If rather than using a sorted dictionary, one uses a tree-based dictionary (e.g., a dictionary implemented using B$^+$-Tree), \system{} synthesizes a tree-based join algorithm.
The tree-based data structures have a significantly better
insertion time in the case where the input data is not already ordered.

\input{figures/indb}

\subsection{Index-Nested Loop Join}
Index-Nested Loop Join can be thought of as a specific case of partitioned join;
when one of the relations is already partitioned (i.e., indexed) on the join key,
there is no need to perform the partitioning in the query processing time.
In this case, the partitioned join operator can be seen as an index-nested loop join operator.

As an example, consider the case where there is a hash-based index, named as !Sind!, for 
relation !S! using the function !part(s.key)!.
The index-nested loop join for !S! and !R! is expressed as follows:

\begin{lstlisting}
let RS = @ht {{ }} in
for(r <- R) {
  for(s <- Sind(part(r.key))) {
    RS( join(r.key, s.key) ) += r.val * s.val
} }
\end{lstlisting} 

\noindent Similarly, one can use a sort-based index (e.g., tree-based or sorted dictionary). In this case, one can again benefit from hinted lookups and inserts by utilizing an iterator. 

\subsection{Group-By Aggregation}
The dictionaries used for computing group-by aggregates can also be hash-based or sort-based.

\subsubsection{Hash-Based Aggregation}
Using a hash table as the underlying data structure for dictionaries, results in 
a hash-based group aggregate.
In our previous group-by aggregate example, the final result is stored in 
the variable !Ragg!, which is instantiated with an empty hash-based dictionary.
For each element !r! of relation !R!, we update the value associated with key !grp(r.key)! with the new value !g(r.key) * r.val! (cf. Figure~\ref{fig:qopt_ha}).

\subsubsection{Sort-Based Aggregation}
We can use a sort-based dictionary for group-by aggregates as well.
Furthermore, if the elements of relation !R! are already sorted based on their group-by key,
one can use hinted inserts, as shown in Figure~\ref{fig:qopt_sa}.

As a result, the group-by aggregate can be computed in linear time rather than $O(n.log(n))$. 
Furthermore, in the case where a sorted result is more preferable (e.g., the presence of !ORDER BY! or if the next operator can benefit from sorted input), using sort-based aggregates 
can be more beneficial over their hash-based counterparts. 

\subsection{Groupjoin Operators}
\label{sec:qopt:gjoin}
Consider an aggregation over the
result of join between two relations.
The aggregate can be interleaved by the join computation. This
is achieved by decomposing the aggregate function into functions
that are only dependent on one of the relations.
Then, one can push the decomposed functions into their corresponding relations. 
Finally, the result of these partial aggregates are joined together.

\subsubsection{Hash-Based Groupjoin}
In essence, this optimization has the effect of fusing a partitioned join operator with an aggregate operator.
In fact, in the specific case of using a hash-table data-structure for dictionaries, this optimization produces a groupjoin operator~\cite{groupjoin}.

\subsubsection{Sort-Based Groupjoin}
In the case of using a sort-based dictionary, \langds{} synthesizes the dual form of
groupjoin operator.
In the previous example, if the elements of !R! are sorted based on !A!,
then one can make the aggregation even faster by having an amortized constant time access for the elements of !Ss! (cf. Figure~\ref{fig:qopt_sgj}).

\subsection{In-DB Learning Engines}
\label{sec:indb}
Recently, there has been an increasing interest in performing machine learning tasks inside a database system.
One of the main techniques for in-DB machine learning is to express the machine learning task as an aggregation query.
This way, one can globally optimize both the feature extraction part of the 
ML task and its training, achieving orders of magnitude performance 
improvement~\cite{Schleich:2019:LAE:3299869.3324961}.

As an example, let us consider a database with two relations:
\col{S}(\textit{s, i, u}),
\col{R}(\textit{s, c}). The goal is to train
a linear regression model that predicts $u$ given features
$\col{F} = \{i, c\}$, where the training dataset is the join of
two relations $\col{Q}=\col{S}\bowtie \col{R}$.

Covariance matrix computation is an essential technique for efficiently training machine learning models such as linear regression~\cite{10.1145/3196959.3196960}. In our example, we consider the part of covariance matrix that considers the interactions of $i$ and $c$, which is defined by  $\Sigma_{x \in \col{Q}}x_\ell * x_j$, where $\ell, j \in \col{F}$.
The na\"ive approach for computing the covariance matrix consists of two stages:
1) computing the join of the input relations using a feature extraction query, followed by 2) aggregations computing the elements of this matrix. 
Thus, this computation can be seen as a multi-aggregate query, the code of which is shown in Figure~\ref{fig:covar_a}.

By interleaving the aggregate and join computations~\cite{Schleich:2019:LAE:3299869.3324961,}, there is no more need to compute the intermediate query !Q!, as can be seen in Figure~\ref{fig:covar_b}. 
Instead the partial aggregates that are dependent only on relation !R!, bound to variable !Ragg!, are computed while scanning this relation.

Next, we introduce a trie index for relation !S!, represented as the nested dictionary !Strie!.
Instead of an iteration over the relation !S!, this program performs a nested iteration
over the trie !Strie! (cf. Figure~\ref{fig:covar_c});
the outer iteration is over the first level of trie !Strie!, and the second iteration
is over the second level !st.val!.

Finally, we further improve the performance by factorizing the independent factors from 
the inner loop and performing loop-invariant code motion to hoist them outside.
The expression !sagg! computes the partial aggregates dependent only on relation !S!.
The last expression computes the final aggregate by multiplying the corresponding
factors from the two partial aggregates of relations !R! and !S! (cf. Figure~\ref{fig:covar_d}).

%% file: figures/indb.tex
\begin{figure*}[t]
\begin{minipage}{0.48\textwidth}
\begin{lstlisting}
let Rp = {{}} in
for(r <- R) {
    Rp( r.key.s ) += {{ r.key -> r.val }}
} ;
let Q = {{}} in
for(s <- S) {
    for(r <- Rp(s.key.s)) {
      Q( {i=s.key.i,c=r.key.c} ) += r.val*s.val
  } } in
let Covar = ref({i_i:double, i_c:double, c_c:double})
for(x <- Q) {
 Covar += { i_i=x.key.i*x.key.i*x.val, 
   i_c=x.key.i*x.key.c*x.val, c_c=x.key.c*x.key.c*x.val }
}
\end{lstlisting}
\subcaption{Initial unoptimized \langds{} expression.}
\label{fig:covar_a}
\vspace{0.5cm}
\end{minipage}~\hspace{0.3cm}~%
\begin{minipage}{0.48\textwidth}
\begin{lstlisting}
let Ragg = {{}} in
for(r <- R) {
    Ragg( r.key.s ) +=
         { m = r.val, c = r.key.c * r.val, 
           c_c = r.key.c * r.key.c * r.val } 
} ;
let Covar = ref({i_i:double, i_c:double, c_c:double}) in
for(s <- S) {
  let r = Ragg(s.key.s) in

  Covar += { i_i=s.key.i*s.key.i*s.val*r.m, 
    i_c=s.key.i*s.val*r.c, c_c=s.val*r.c_c }
}
\end{lstlisting}
\subcaption{After interleaving join and aggregations.}
\label{fig:covar_b}
\vspace{0.5cm}
\end{minipage}

\begin{minipage}{0.48\textwidth}
\begin{lstlisting}
let Ragg = {{}}
for(r <- R) {
    Ragg( r.key.s ) += { m = r.val, c = r.key.c * r.val, 
                     c_c = r.key.c * r.key.c * r.val }
} in
let Covar = ref({i_i:double, i_c:double, c_c:double}) in
for(st <- Strie) {
 let r = Ragg(st.key) in

 for(s <- st.val) {

  Covar += { i_i=s.key.i*s.key.i*s.val*r.m, 
    i_c=s.key.i*s.val*r.c, c_c=s.val*r.c_c }

 } 
}
\end{lstlisting}
\subcaption{After introducing trie indices.}
\label{fig:covar_c}
\end{minipage}~\hspace{0.3cm}~%
\begin{minipage}{0.48\textwidth}
\begin{lstlisting}
let Ragg = {{}}
for(r <- R) {
    Ragg( r.key.s ) += { m = r.val, c = r.key.c * r.val, 
                    c_c = r.key.c * r.key.c * r.val }
  } in
let Covar = ref({i_i:double, i_c:double, c_c:double}) in
for(st <- Strie) {
 let r = Ragg(st.key) in
 let sagg = ref({i_i:double, i:double, m:double}) in
 for(s <- st.val) {
   sagg += { i_i=s.key.i*s.key.i*s.val, 
     i=s.key.i*s.val, m=s.val}
 } ;
 Covar += { i_i=sagg.i_i*r.m, 
   i_c=sagg.i*r.c, c_c=sagg.m*r.c_c }
}
\end{lstlisting}
\subcaption{After factorization and loop-invariant code motion.}
\label{fig:covar_d}
\end{minipage}
\vspace{-0.2cm}
\caption{Different \langds{} expressions representing covariance matrix computation over join of two relations.}
\label{fig_covar}
\end{figure*}

%% file: cost.tex
\input{figures/costinfer}

\section{Cost Engine}
\label{sec:cost}
In this section, we present how \system automatically infers the execution cost of \langds{} programs by combining machine learning and program reasoning.
First, we present the dictionary cost model trained by regression models in Section~\ref{sec:cost:dict}.
Then, we show the cost inference rules required for estimating the run-time cost of \langds{} programs using the dictionary cost model and cardinality model in Section~\ref{sec:costinfer}. 

\subsection{Dictionary Cost Model: Regression Learning}
\label{sec:cost:dict}

The training set is generated by profiling the run time of the insert and lookup operations for different
dictionary implementations. 
For insert, 
the profiling is generated for different dictionary sizes. 
For lookup, the profiling is generated 
for different dictionary sizes and number of accessed tuples. 
Note that the lookup operator shows a different runtime behavior depending on whether the 
lookup is successful or not. Hence, the training set contains profiling for both 
successful and unsuccessful lookups. 

Another important feature is the orderedness of the input data.
As observed in Section~\ref{sec:langds}, the sort-based dictionaries can use hinted 
lookups and inserts when the input keys is ordered.
We profile the situations where the sequence of key-value pairs to insert or to look up are ordered or unordered.
Note that the performance of hash tables is independent of orderedness of keys.

In summary, our training dataset has the following features: dictionary size, number of accessed tuples, and orderedness of data.
We also enhanced the features by adding the logarithm of dictionary size and number of accessed tuples as explained in Section~\ref{sec:exp:cost}.
The labels are the run time performance for various operations in milli-seconds.
As future work, one can investigate further features, such as data distribution parameters, tuple arity, and data type.


The next step involves training a regression model over the profiling training set
to predict the run time cost of dictionary operations.
This defines our dictionary cost model.
We have used a wide range of regression models, the behavior of which can be observed in Section~\ref{sec:exp:cost}.

Next, we use program reasoning to derive the cost of \langds{} programs.

\subsection{\langds{} Cost Model: Program Reasoning}
\label{sec:costinfer} 
The trained regression models give a cost estimate for individual dictionary operations.
However, the program synthesizer requires the cost estimate for \langds{} expressions.
Figure~\ref{fig:costinfer} shows how the dictionary cost model ($\Delta$ \regrmodelshape) and cardinality model ($\Sigma$ \cardmodelshape) are combined with the runtime context of \langds{} expressions to
derive their execution cost.

In Figure~\ref{fig:costinfer}, the top and bottom parts of the inference rules specify the premises and the conclusions, respectively. 
As an example, the premises for the second inference rule specifies that for the expression !for(x <- e1) e2!, the runtime context of !e2! ($\Gamma'$) should extend the runtime context of !e1! ($\Gamma$) by recording that the number of invocations is multiplied by the cardinality of !e1! ($\Sigma_{card}(\texttt{e1})$).
The conclusion of this rule specifies that the execution cost of the mentioned expression is $c_1+c_2$ given that the execution cost of !e1! and !e2! is $c_1$ and $c_2$, respectively.

\smartpara{Example for Cost Inference}
To give a more concrete example, consider the hash-based group-by aggregate from Figure~\ref{fig:qopt_ha}. The initial runtime context is set as $\Gamma=(\Gamma_{calls}=1,\Gamma_{cond}=1)$, specifying that the number of invocations and the probability of taking the execution path are both $1$. The runtime context of the update statement, which is the body of the loop, is modified with $\Gamma'_{calls}=\Sigma_{card}(\texttt{R})$. In the case of having a filter, $\Gamma'_{cond}$ is modified according to the inference rule of !if! statements. 

The execution cost of the update statement is computed following the corresponding inference rule. First, assume that the cost for the group-by key (!grp(r.key)!) and value (!agg(r.key) * r.val!) are $c_k$ and $c_v$, respectively. Second, the number of invocations of the update statement and the size of dictionary are computed as
$C=\Gamma'_{calls}*\Gamma'_{cond}=\Sigma_{card}(\texttt{R})$ and
$N=\Sigma_{card}(\texttt{Ragg})$. Then, the number of lookup hits is computed as $H=C-N=\Sigma_{card}(\texttt{R})-\Sigma_{card}(\texttt{Ragg})$. 
Finally, total execution time is computed as $c_k+c_v+\Delta_{lus}(H,N)+\Delta_{luf}(N,N)+\Delta_{ins}(N)$, where $\Delta_{lus}(H,N)$ is the estimation cost returned by the regression model for successful lookups of $H$ elements in a dictionary of size $N$, $\Delta_{luf}(N,N)$ specifies the cost model for failed lookups of $N$ elements in a dictionary of size $N$, and $\Delta_{ins}(N)$ specifies the cost for inserting $N$ elements to a dictionary.

In the next section, we show how this cost model can be used for program synthesis.

%% file: figures/costinfer.tex
\begin{figure*}[t]
\centering
\begin{tabular}{c}
$\Sigma,\Delta,\Gamma \vdash$ !e1!: $c_1$ $\quad$ $\Sigma,\Delta,\Gamma \vdash$ !e2!: $c_2$
\\\hline
$\Sigma,\Delta,\Gamma \vdash$ !e1; e2!: $c_1+c_2$
\end{tabular}\hspace{0.4cm}
\begin{tabular}{c}
$\Sigma,\Delta,\Gamma \vdash$ !e1!: $c_1$ $\quad$ $\Sigma,\Delta,\Gamma' \vdash$ !e2!: $c_2$
$\quad$ $\Gamma'=\Gamma[\Gamma'_{calls} = \Gamma_{calls} * \Sigma_{card}(\texttt{e1})]$
\\\hline
$\Sigma,\Delta,\Gamma \vdash$ !for(x <- e1) e2!: $c_1+c_2$
\end{tabular}

\vspace{0.25cm}

\begin{tabular}{c}
$\Sigma,\Delta,\Gamma \vdash$ !e1!: $c_1$ $\quad$ $\Sigma,\Delta,\Gamma' \vdash$ !e2!: $c_2$ $\quad$ $\Sigma,\Delta,\Gamma'' \vdash$ !e3!: $c_3$
$\quad$ $\Gamma'=\Gamma[\Gamma'_{cond} = \Gamma_{cond} * \Sigma_{sel}(\texttt{e1})]$
$\quad$ $\Gamma''=\Gamma[\Gamma''_{cond} = \Gamma_{cond} * (1-\Sigma_{sel}(\texttt{e1}))]$
\\\hline
$\Sigma,\Delta,\Gamma \vdash$ !if(e1) then e2 else e3!: $c_1+c_2+c_3$
\end{tabular}

\vspace{0.25cm}

\begin{tabular}{c}
$\Sigma,\Delta,\Gamma \vdash$ !e1!: $c_1$ $\quad$ $\Sigma,\Delta,\Gamma \vdash$ !e2!: $c_2$
$\quad$ $C=\Gamma_{calls} * \Gamma_{cond}$
$\quad$ $N=\Sigma_{card}(\texttt{e1})$
$\quad$ $\sigma=\Sigma_{dist}(\texttt{e2})/N$
$~~~$ $H=\sigma * C$
$~~~$ $M=C-H$
$~~~$ $ds=\Gamma_{dict}(\texttt{e1})$
\\\hline
$\Sigma,\Delta,\Gamma \vdash$ !e1(e2)!: $c_1+c_2+\Delta_{lus}(ds,H,N)+\Delta_{luf}(ds,M,N)$
\end{tabular}

\vspace{0.25cm}

\begin{tabular}{c}
$\Sigma,\Delta,\Gamma \vdash$ !e1!: $c_1$ 
$\quad$ $\Sigma,\Delta,\Gamma \vdash$ !e2!: $c_2$
$\quad$ $\Sigma,\Delta,\Gamma \vdash$ !e3!: $c_3$
$\quad$ $C=\Gamma_{calls} * \Gamma_{cond}$
$\quad$ $N=\Sigma_{card}(\texttt{e1})$
$\quad$ $H=C-N$
$\quad$ $ds=\Gamma_{dict}(\texttt{e1})$
\\\hline
$\Sigma,\Delta,\Gamma \vdash$ !e1(e2) += e3!: $c_1+c_2+c_3+\Delta_{lus}(ds,H,N)+\Delta_{luf}(ds,N,N)+\Delta_{ins}(ds,N)$
\end{tabular}

\vspace{0.5cm}

\begin{tabular}{|l||c|c||c|c||c|c|}
\hline
$\Sigma$ \cardmodelshape &
$\Sigma_{card}$ & Cardinality of the given dictionary &
$\Sigma_{dist}$ & Number of distinct elements &
$\Sigma_{sel}$ & Selectivity of condition \\ \hline
$\Delta$ \regrmodelshape&
$\Delta_{lus}$ & Cost of successful lookup &
$\Delta_{luf}$ & Cost of failed lookup &
$\Delta_{ins}$ & Cost of insertion \\ \hline
$\Gamma$&
$\Gamma_{calls}$ & Total number of invocations &
$\Gamma_{cond}$ & Accumulative probability &
$\Gamma_{dict}$ & Dictionary implementation \\ \hline
\end{tabular}
\vspace{0.1cm}
\caption{Cost inference of a subset of \langds{} expressions. The contexts used in the inference rules are as follows: $\Sigma$ corresponds to cardinality model, $\Delta$ corresponds to the dictionary cost model, and $\Gamma$ corresponds to the runtime context.}
\label{fig:costinfer}
\end{figure*}

%% file: synthesis.tex
\section{Program Synthesis}
\label{sec:ds_synthesis}

This section presents fine-tuning of the dictionary implementations by using program synthesis.
The input to the program synthesis is an \langds{} expression for which the join order is already specified.
Then, the search space for using different combinations of dictionary implementations is generated.
By using the cost inference engine shown in the previous section, we find the \langds{} with dictionary implementations that lead to the lowest execution time.

Algorithm~\ref{alg:ds_sel} shows a greedy algorithm for the data-structure selection process.
First, the distinct dictionaries that exist in the input \langds{} expression are extracted (Line 2).
Then, a dependency graph~\cite{pdg} among these dictionaries is created, so that we traverse them in dependency order (Line 3).

For each dictionary symbol (\texttt{sym}), we select the data structure with the minimum total run-time cost estimate (Line 6). The function \textsc{Cost} uses the inference rules presented in Figure~\ref{fig:costinfer}, and its runtime context is updated to use $ds$ for the dictionary symbol \texttt{sym}. Accordingly we update the runtime context with best dictionary implementation for \texttt{sym} (Line 7).

Finally, we replace the dictionary symbols in \texttt{exp} by the implementation choices collected in the runtime context (Line 9).
This is achieved by changing the annotations !@ds! in the statements where the dictionary symbols are introduced with the corresponding dictionary implementation.

We observe that for many analytical queries, where one uses pipelining or the intermediate results only used for probing, there is no dependency between dictionary symbols. In such cases, the greedy algorithm finds the optimal program (assuming that the cost/cardinality models are precise).
However, in cases where one needs to iterate over the intermediate dictionaries (e.g., in-DB ML and TPCH query 18), the greedy algorithm can fall into local optimum. 
We leave the usage of further search algorithms~\cite{yan2019generating,trummer2017solving,ioannidis1990randomized} for future work.

\begin{algorithm}[t!]
\begin{flushleft}
\begin{tabular}{l}
\textbf{Inputs:}\\
\texttt{exp}: Input expression\\
$\Sigma$: Cardinality model \cardmodelshape\\
$\Delta$: Dictionary cost model \regrmodelshape\\
$\alglist{DS}$: Dictionary implementations \dictimplshape\\
\end{tabular}
\end{flushleft}

\begin{algorithmic}[1]
\Function{ProgramSynthesis}{\texttt{exp}, $\Sigma$, $\Delta$, $\alglist{DS}$}
\State $\alglist{Dict}$ $\gets$ \Call{ExtractDictSymbols}{\texttt{exp}}
\State $\alglist{DAG}$ $\gets$ \Call{DependencyGraph}{\texttt{exp}, $\alglist{Dict}$}
\State $\Gamma \gets (\Gamma_{calls}=1, \Gamma_{cond}=1)$
\For{\texttt{sym} $\gets \alglist{DAG}$}
\State $ds_{best} \gets \argmin\limits_{ds \in \alglist{DS}}$ \Call{Cost}{\texttt{exp}, $\Sigma$, $\Delta$, $\Gamma[\Gamma_{dict}(\texttt{sym})=ds]$} %
\State $\Gamma \gets \Gamma[\Gamma_{dict}(\texttt{sym})=ds_{best}]$
\EndFor
\State \texttt{final} $\gets$ \Call{ChooseDictDS}{\texttt{exp}, $\Gamma_{dict}$}
\State \Return{\texttt{final}}
\EndFunction
\end{algorithmic}
\vspace{0.2cm}
\caption{A greedy algorithm for program synthesis.}
\label{alg:ds_sel}
\end{algorithm}

%% file: exp.tex
\section{Experimental Results}
\label{sec:exp}
In this section, we investigate the experimental results of \system. Our findings are summarized as follows:

\begin{itemize}[leftmargin=*]
\setlength\itemsep{0em}
\item The predicted cost by the dictionary cost model is proportional to the actual time spent for each operation using different dictionary implementations (Section~\ref{sec:exp:cost:dict}). 
\item Utilizing the \langds{} cost model to find the best query operator can prevent a slowdown compared to the best plan in most cases (Section~\ref{sec:exp:cost:prog}). 
\item If we resort to a single dictionary implementation for the entire query, we observe our engine performs on par with state-of-the-art analytical engines (Section~\ref{sec:exp:tpch}).
\item By using several dictionary implementations for one query, we observe an average of 70\% performance improvement over the version that uses a single dictionary implementation.
This is in particular the case for queries that can benefit from sort-based group-by and join (Section~\ref{sec:exp:tpch}).
\item Overall, \system outperforms the state-of-the-art query engines Typer and Tectorwise by 1.5x and respectively 2x on average (Section~\ref{sec:exp:tpch}), while also recovering the runtime performance of the LMFAO specialized in in-DB machine learning (Section~\ref{sec:exp:indbml}).
\end{itemize}

\subsection{Experimental Setup}
All experiments are performed on two machines:
\begin{itemize}[leftmargin=*]
\setlength\itemsep{0em}
\item \textbf{Machine 1} is an iMac equipped with
an Intel Core i5 CPU running at 2.7GHz, 32GB of DDR3 RAM with OS X 10.13.6. We use CLang 900.0.39.2 for compiling the generated C++ code using the O3 optimization flag.
\item \textbf{Machine 2} is equipped with
Intel(R) Core(TM) i7-4770 at 3.40GHz, 32GB of DDR3 RAM, running Ubuntu
18.04. We use g++ 6.4.0 for compiling the generated C++ code using the O3
flag.  
\end{itemize}

\noindent
We use the following dictionary implementations:

\begin{itemize}[leftmargin=*]
\setlength\itemsep{0em}
\item \dsdict{\stlumap{}}: C++ STL hashing.
\item \dsdict{\rbdict{}}: robin-hood hashing~\cite{rbdict}.
\item \dsdict{\tsldict{}}: hopscotch hashing~\cite{tsldict}.
\item \dsdict{\boostumap{}}: Boost hashing~\cite{boostumap}.
\item \dsdict{\stlmap{}}: C++ STL Red-Black tree dictionary.
\item \dsdict{\boostfmap{}}: Boost sorted flat array~\cite{boostumap}.
\item \dsdict{\tlxdict{}}: TLX $\text{B}^+$-tree dictionary~\cite{tlxdict}.
\item \dsdict{\absldict{}}: Abseil B-tree dictionary~\cite{absldict}.
\end{itemize}

{\bf\noindent Competitors.} We benchmark our engine against the following in-memory engines\footnote{Comparing against disk-based engines requires our cost model to consider hardware characteristics of HDDs and SDDs (e.g., erase time), which we leave for future.}: (1) Typer and (2) Tectorwise [23], the open source implementation\footnote{https://github.com/TimoKersten/db-engine-paradigms} of HyPer~\cite{Neumann11} and Vectorwise~\cite{monetdb-handwritten},
and (3) the in-DB ML engine LMFAO~\cite{Schleich:2019:LAE:3299869.3324961}.\footnote{https://github.com/fdbresearch/LMFAO}
Typer and LMFAO use query compilation and Tectorwise uses vectorization. The code for queries in all systems is in C++.\footnote{Generating LLVM or machine code can improve query compilation time~\cite{kohn2018adaptive}. However, improving compilation time is beyond the scope of this paper; instead we only focus on query execution time.}

For all the experiments, we compute the average of ten subsequent runs.  
We perform all experiments using a single core. We leave the experimentation for multi-core environments for the future since it requires dealing with parallelization concerns (e.g., lock-based vs. lock-free data structures).
The loading time of the database into main memory is not considered.
For in-DB machine learning experiments, all relations are 
sorted by their join attributes for both \system{} and LMFAO.

\subsection{Cost Engine Performance}
\label{sec:exp:cost}
\subsubsection{Dictionary Cost Model}
\label{sec:exp:cost:dict}
In this section, we report the performance of our dictionary cost model using various regression models for prediction. Several regression models were trained over the profiling training set to predict the run time cost of dictionary operations using three different methods:
\begin{itemize}[leftmargin=*]
	\item All in One Model: A single regression model is used for cost prediction. The model uses the dictionary size, number of accessed tuples, orderedness of data, dictionary type, and the operation as features. The last two mentioned features are passed to the model in the one-hot encoded format. 
	\item Individual Models Without Feature Engineering: 32 different regression models are constructed based on the combinations of data order, dictionary implementation, and operations. Each model takes the dictionary size and the number of accessed tuples as features.
	\item Individual Models With Feature Engineering: It is constructed in the same way to the previous method. However, its features are enriched with the logarithm values of the dictionary size and the number of accessed tuples.
\end{itemize}

\begin{figure}[t]
	\includegraphics[width=\columnwidth]{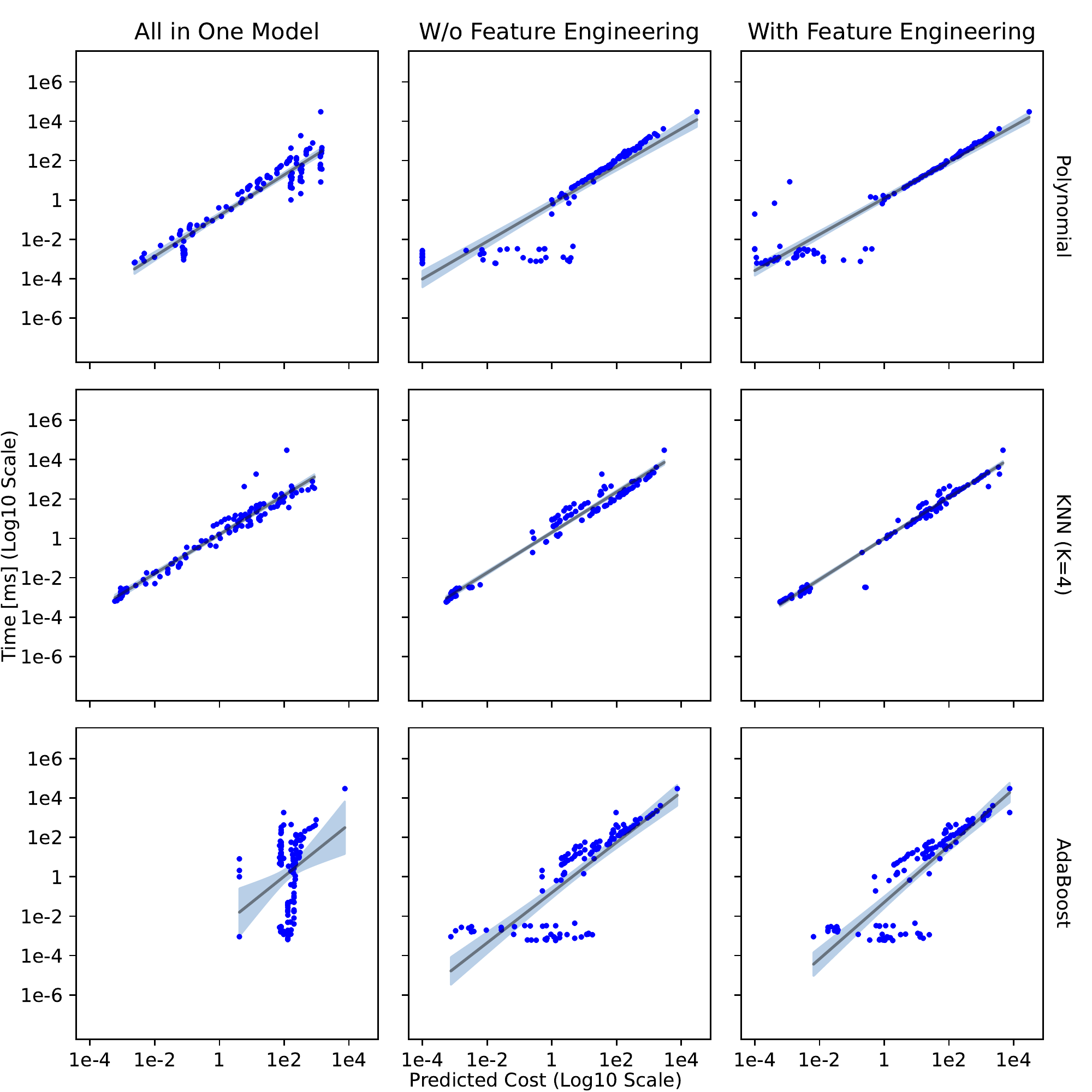}
	\caption{Comparison of the prediction of different regression models trained under various methods with operation running times.}
	\label{fig:cost_pred_final}
\end{figure}

Figure~\ref{fig:cost_pred_final} shows a comparison between our cost model with the actual run-time spent on basic operations (lookup and insert). We observe that in most cases, our cost model is proportional to the actual time on a logarithmic scale. Since the logarithm of dictionary size accurately captures the relationship between dictionary size and actual operation cost, the models that have been trained with feature engineering outperform other methods. Overall, KNN with $K=4$ and trained with logarithmic features performs the best among all these models. 

\subsubsection{\langds{} Cost Model}
\label{sec:exp:cost:prog}
\begin{figure}[t]
\vspace{-0.9cm}
	\includegraphics[width=\columnwidth]{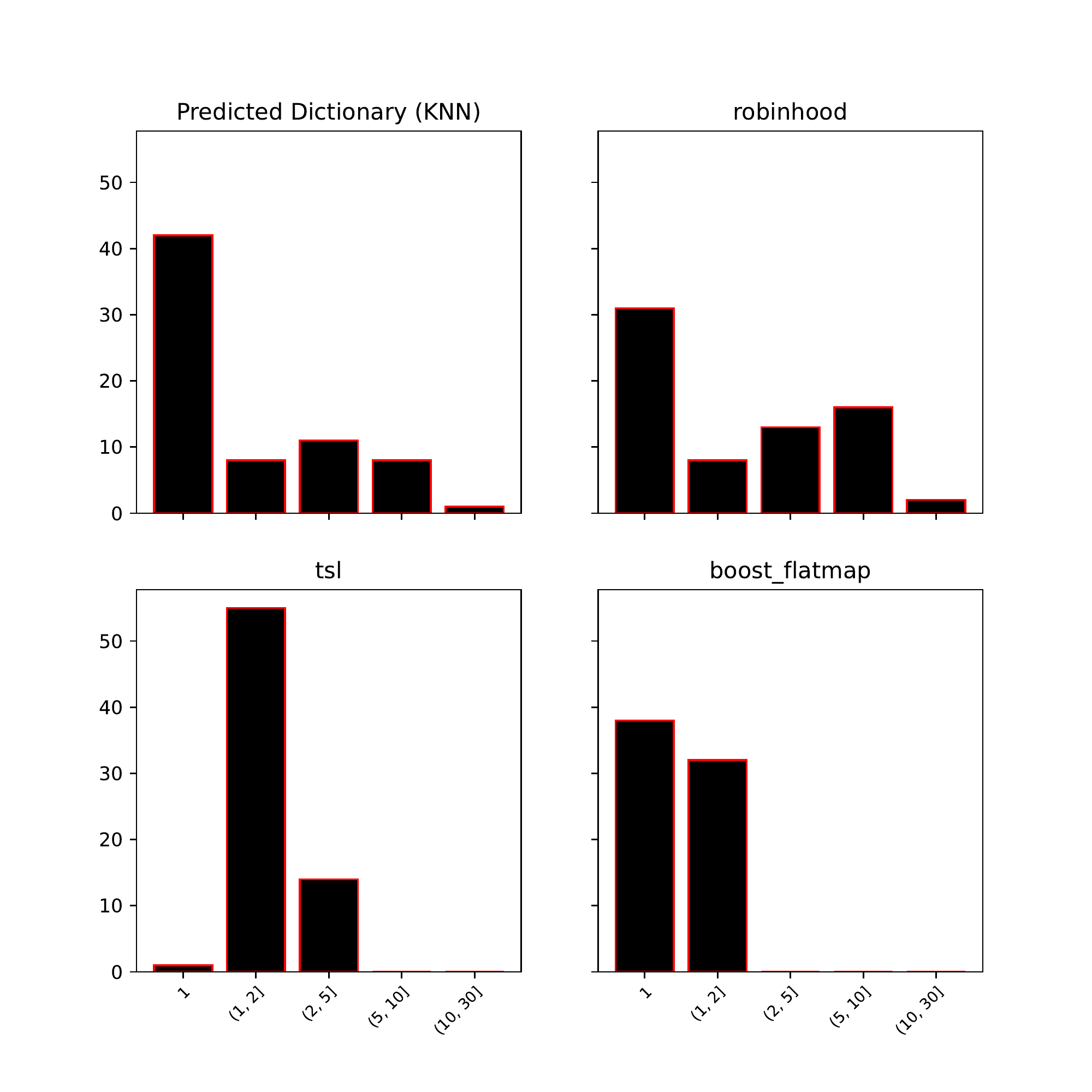}
\vspace{-0.9cm}
	\caption{Slowdowns compared to the best option for the group-by operation.}
	\label{fig:slowdown}
\end{figure}

To evaluate the \langds{} cost model, we benchmark the \langds{} program for the group-by operation by varying the selectivity of the input relation. Overall, we generate 70 different experiments by logarithmically increasing the selectivity. Figure~\ref{fig:slowdown} demonstrates that our cost model mostly suggests the best dictionary choice based on given features. Selecting the best dictionary based on our cost model's prediction outperforms each of the implementations individually.

\subsection{Analytical Query Engines} 
\label{sec:exp:tpch}
In this section, we investigate the performance of \system for OLAP workloads.
For this purpose, we use a representative subset of TPCH queries involving joins and 
aggregations with a wide range of intermediate cardinalities~\cite{kersten2018everything,Boncz2014}.
We compare the performance of generated optimized code for two best hash-based dictionaries,
the best sort-based dictionary, the fine-tuned versions (M1 Tuned and M2 Tuned), Typer, and Tectorwise.

\begin{figure}[t]
\begin{tabular}{m{0.5em}|c}
\rotatebox{90}{\textbf{Machine 1}} &
\begin{subfigure}{0.9\columnwidth}
\includegraphics[width=\columnwidth]{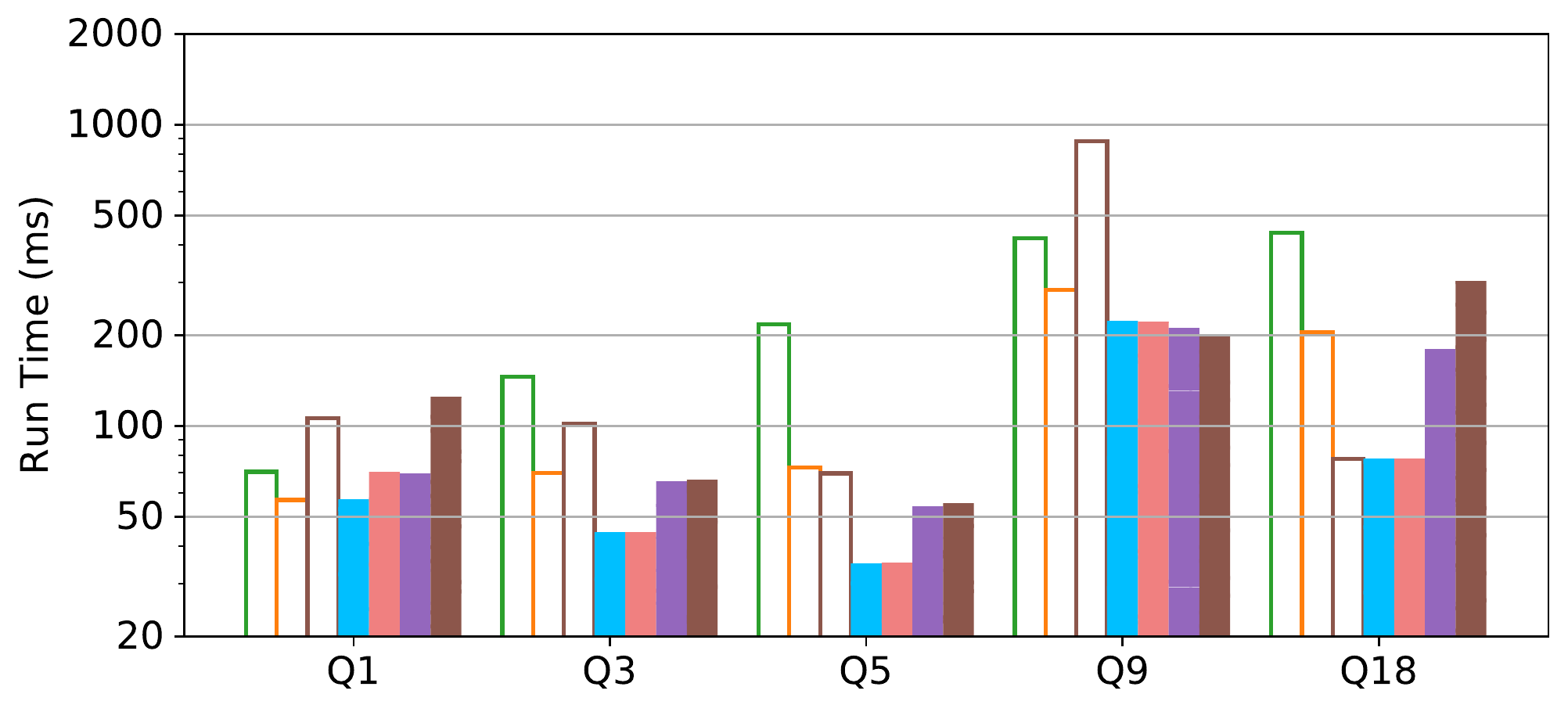}
\end{subfigure}
\\ \hline
\rotatebox{90}{\textbf{Machine 2}} &
\begin{subfigure}{0.9\columnwidth}
\includegraphics[width=\columnwidth]{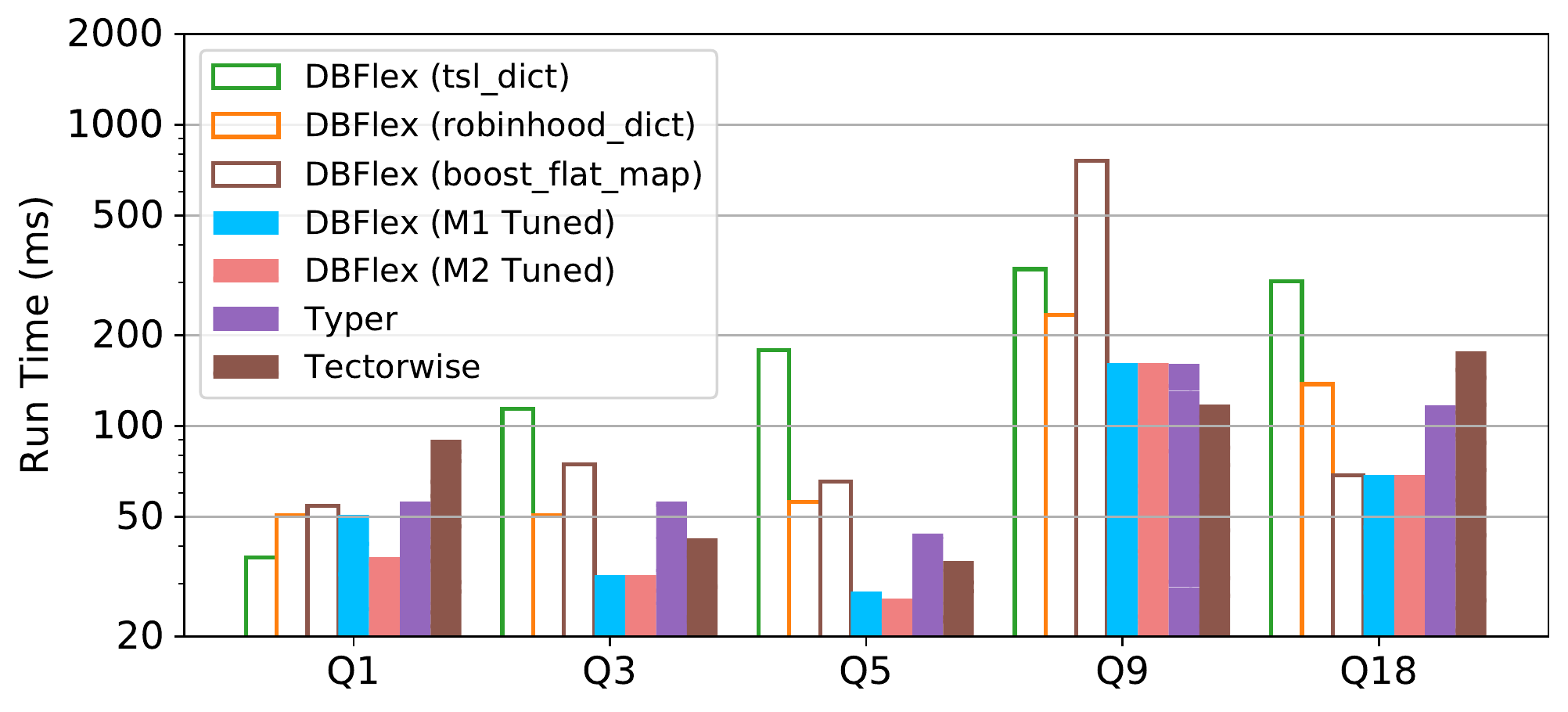}
\end{subfigure}
\\
\end{tabular}
\caption{Run time results for TPCH queries comparing different strategies employed for choosing data-structures
in \system and state-of-the-art query processing engines.}
\label{fig:exp:tpch}
\end{figure}

As Figure~\ref{fig:exp:tpch} shows, we observe that overall the fine-tuned optimized queries
perform better or the same as the Typer and Tectorwise engines.
Furthermore, we observe that in most cases, the tuned versions for the two machines produce identical query plans.
In Q1, the only involved dictionary favors a \dsdict{\rbdict{}} dictionary in machine 1, instead of \dsdict{\tsldict{}} in machine 2.

Q3, Q5, and Q9 all involve multiple joins, and for all of them the hash-based \dsdict{\rbdict{}} 
dictionaries show promising performance.
However, the fine-tuned optimized query for all of them involves a mixture of \dsdict{\boostfmap{}} and \dsdict{\rbdict{}}.
Furthermore, all these queries show good performance for vectorized engines such as Tectorwise.
Especially, Q9 involves a large intermediate dictionary, for which a vectorized engine is better at hiding memory stalls for a large
intermediate hash join~\cite{kersten2018everything}.

Finally, Q18 involves a high-cardinality aggregation operator. For this query,
we observe that sort-based dictionaries such as \dsdict{\boostfmap{}} outperform hash-based ones.
A particular interesting characteristic of this query is that two instances of the
sort-based dictionaries cannot use the hinted version of lookup. Nevertheless, due to the low 
cardinality of the corresponding intermediate dictionaries, makes the overall non-hinted and logarithmic 
lookup computation time of sort-based dictionaries is comparable to the constant lookup time of hash-based ones. 
Thus, the overall performance of sort-based dictionaries is better than hash tables.

\subsection{In-DB Machine Learning}
\label{sec:exp:indbml}
As the final set of experiments, we show the performance of \system for in-DB ML workloads. We use two real-world datasets: 1)
\emph{Favorita}~\cite{favorita}, which is a publicly available Kaggle dataset, and 2)
\emph{Retailer} is a dataset from a US retailer~\cite{Schleich:2016:LLR:2882903.2882939}.
Both datasets are used in retail forecasting scenarios and have a snowflake schema with 4 
dimension tables for both and with
fact tables of 87 million and 125 million tuples, respectively. We only use the continuous 
attributes of these datasets, which are 6 and 35 attributes, respectively.

\begin{figure}[t]
\begin{tabular}{m{0.5em}|c}
\rotatebox{90}{\textbf{Machine 1}} &
\begin{subfigure}{0.9\columnwidth}
\includegraphics[width=\columnwidth]{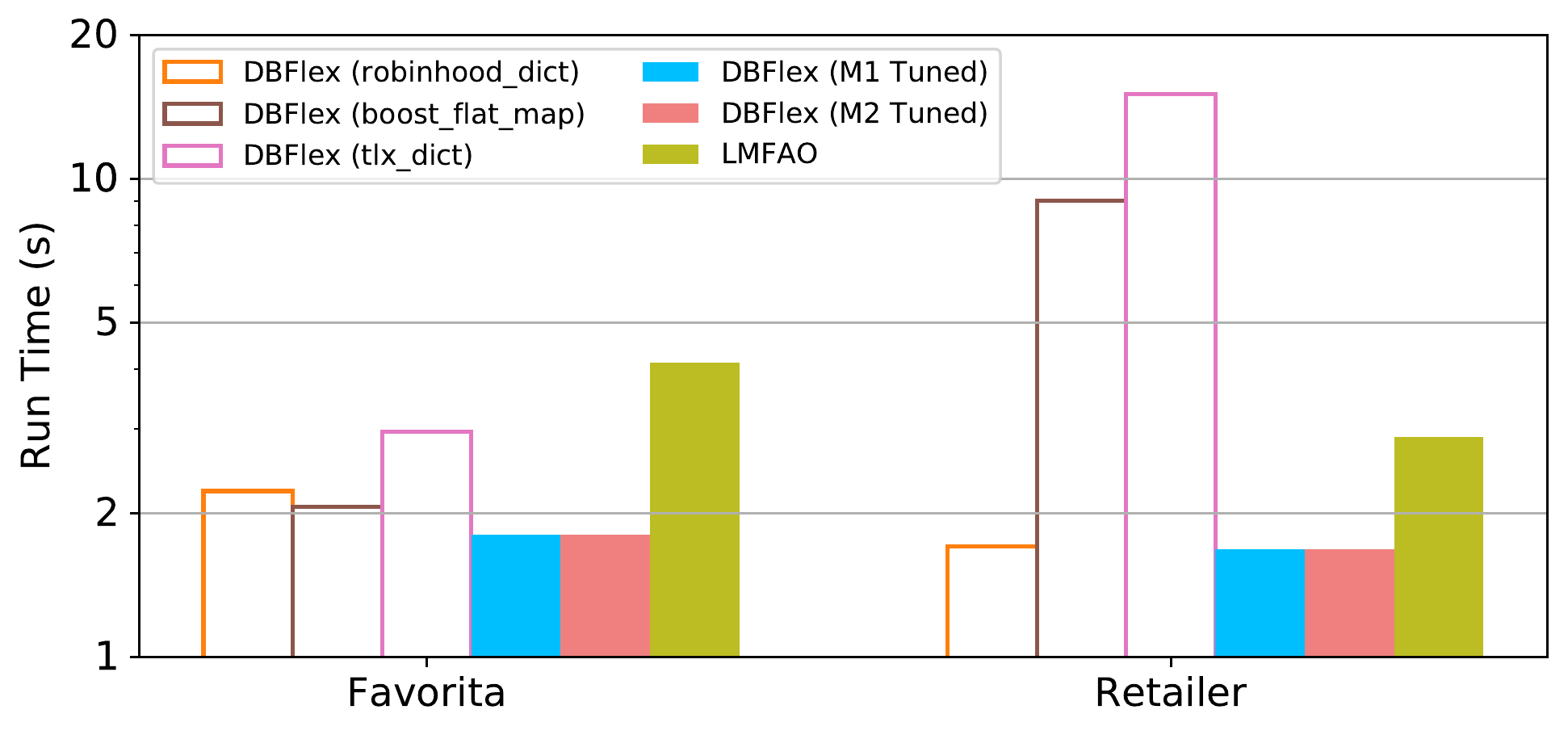}
\end{subfigure}
\\ \hline
\rotatebox{90}{\textbf{Machine 2}} &
\begin{subfigure}{0.9\columnwidth}
\includegraphics[width=\columnwidth]{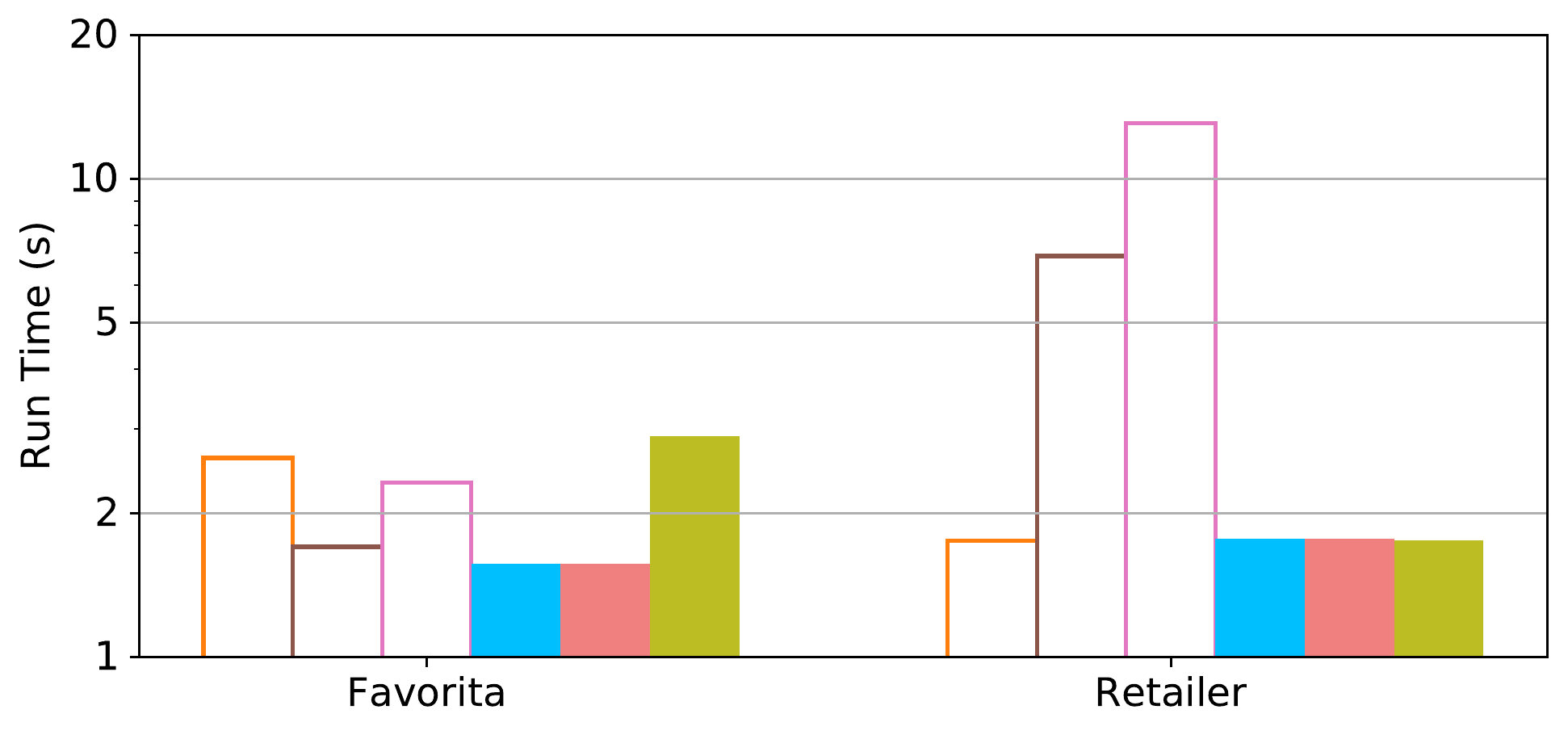}
\end{subfigure}
\\
\end{tabular}
\caption{Run time results for in-DB ML of the covariance matrix computation comparing different strategies employed for choosing data-structures
in \system and LMFAO.}
\label{fig:exp:indb}
\end{figure}

Figure~\ref{fig:exp:indb} shows the run-time comparison of different configurations of 
\system{} for computing the covariance matrix on these two datasets.
As the input relations are already indexed as ordered tries, sort-based dictionaries show better performance.
Thus, we compare the generated code using two best sort-based dictionaries and the best
hash table.

For the Favorita dataset, we observe that \dsdict{\boostfmap{}} outperforms \dsdict{\rbdict{}} and \dsdict{\tlxdict{}}. 
The sort-based dictionaries use hinted lookups and inserts in all cases by default thanks
the ordered nature of their input data.
However, the fine-tuned version for both machines prefer a non-hinted lookup in the case 
where the size of the intermediate dictionary is too small and there are too many 
failed lookups in deeply nested loops. This kind of knowledge is not possible to be 
captured by the competitor systems such as LMFAO.

The Retailer dataset shows better performance for hash-based dictionaries. 
This is because of the failed lookups in deeply nested loops, which makes 
hinted lookups perform worse. The intermediate dictionaries in the fine-tuned generated code are a mixture of \dsdict{\rbdict{}} 
and \dsdict{\boostfmap{}}, and one of the lookups of \dsdict{\boostfmap{}} is non-hinted.
This makes the performance of fine-tuned \system{} better than LMFAO in machine 1, and
comparable to it in machine 2.

%% file: related.tex
\section{Related Work}
\label{sec:related}

\smartpara{Query Languages}
Nested relational model~\cite{roth1988extended}, monad calculus, and monoid comprehension~\cite{monad-calc-1, monad-calc-2, monad-comprehension, query-comprehension, query-comprehension-2, monoid-comprehension, wong2000kleisli, Buneman:1995:PPC:210500.210501} are query languages for nested collections, whereas AGCA~\cite{dbtoaster}, FAQ~\cite{abo2016faq}, and HoTTSQL~\cite{chu2017hottsql} 
represent relations as bags.
The dictionary-oriented nature of \langds{} combines these two lines of work; both relations and group-by aggregates are represented as dictionaries. 
Furthermore, \langds{} allows hash-based and sort-based data structures to be used 
for dictionaries, with the capability of encoding hinted lookup and insertions for sort-based ones.

\smartpara{Query Compilation} 
Just-in-time compilation of queries allows for generating specialized engines and has been heavily used for analytical query processing~\cite{krikellas, 
Neumann11, dbtoaster, legobase_tods, dblablb, DBLP:journals/debu/ViglasBN14, crotty2015tupleware, Nagel:2014:CGE:2732977.2732984, karpathiotakis2015just, spark-sql,palkar2017weld}.
In parallel, the compilers of functional languages heavily investigated the specialized low-level code generation with focus on fusion of intermediate collections~\cite{Kiselyov:2017:SFC:3009837.3009880,Mainland:2013:EVI:2500365.2500601,jfppushpull}.
Recently, there has been several efforts on low-level query plan languages, mainly inspired by functional collections~\cite{DBLP:journals/pvldb/Fent021,10.1145/3448016.3457288,DBLP:journals/pvldb/BandleG21,voodoo}.
None of these systems have focused on fine-tuning data structures and do not support automatic inference of cost models. 

Data-structure specialization in LegoBase~\cite{legobase_tods} and LB2~\cite{tahboub2018architect} focuses on
more aggressive partial evaluation for the provided hash-table implementations, without
tuning based on cardinalities or using sort-based dictionaries.

Chestnut~\cite{yan2019generating} uses integer linear programming to specify data layouts used for database-backed applications, using manually specified cost models.
Micro adaptivity~\cite{ruaducanu2013micro} is a technique for choosing
the best function implementation in runtime for Vectorwise.
Similar to \system{}, it frees database developers from manually writing cost models.
\system{} generalizes this idea to higher-level decisions such as the choice of data structures, while combining it with query compilation.

\smartpara{In-DB Machine Learning}
Training ML models inside the database system by avoiding the materialization of join has recently gained great interest in the community.
The current solutions are currently divided into two categories. 
First, systems such as Morpheus~\cite{chen2017towards,li2019enabling} cast the in-DB ML task as a linear algebra problem. 
For example, such tasks are expressed on top of linear algebra libraries of R~\cite{chen2017towards} and NumPy~\cite{li2019enabling}. 
The second category casts the in-DB ML task as a multi-aggregate analytical query. Systems such as F~\cite{Olteanu:2016:FD:3003665.3003667, Schleich:2016:LLR:2882903.2882939}, AC/DC~\cite{Khamis:2018:AIL:3209889.3209896}, LMFAO~\cite{Schleich:2019:LAE:3299869.3324961}, IFAQ~\cite{ifaq-cgo,ifaq_ir}, SDQL~\cite{shaikhha2021functional}, as well as \system fall into this category.
None of the mentioned systems support fine-tuning hash-based and sort-based data structures as well as automatic inferring of cost models.

\smartpara{Cost Inference}
The run-time cost estimation of programs has been heavily investigated for databases~\cite{cowbook} and programming languages~\cite{jost2010static,hoffmann2011multivariate,albert2009termination,wang2017timl}.
In addition to improving the performance, cost estimation can be used for verification purposes (e.g., ensuring resource usage is bounded for embedded devices)~\cite{hoffmann2017towards}.
The cost model used in \system estimates the run-time cost by relying on the trained models over actual profile data, as opposed to the alternative approaches which mostly rely on asymptotic reasoning~\cite{wang2017timl}. 

\smartpara{Auto-Tuning}
Automatic tuning of performance-critical kernels is a well-investigated topic in
the high-perfor\-mance computing and compilers communities~\cite{ashouri2018survey}. Examples include 
FFTW~\cite{frigo1998fftw} and Spiral~\cite{spiral} for Fourier transforms, and LGen~\cite{Spampinato:2014:BLA:2581122.2544155} and ATLAS~\cite{whaley2001automated} for linear algebra.
None of the mentioned systems support query processing workloads, and do not fine-tune data structures.

%% file: appendix.tex
\section{Micro Benchmarks}
\label{sec:exp:micro}
We next report on micro-benchmarks for the eight dictionary implementations in the following disciplines: (1) inserts of a varying number of data points into a dictionary; (2) successful and (3) unsuccessful lookups for a varying number of keys into
dictionaries of varying size.
The key-value pairs to be inserted, or the keys to be looked up
are integer values generated following a uniform distribution.
The keys can be either ordered or unordered.

\begin{figure}[h]
\begin{tabular}{m{0.5em}|c|c}
& \textit{\small Unordered} & \textit{\small Ordered} \\ \hline
\rotatebox{90}{\textbf{Machine 1}} &
\begin{subfigure}{0.42\columnwidth}
\includegraphics[width=\columnwidth]{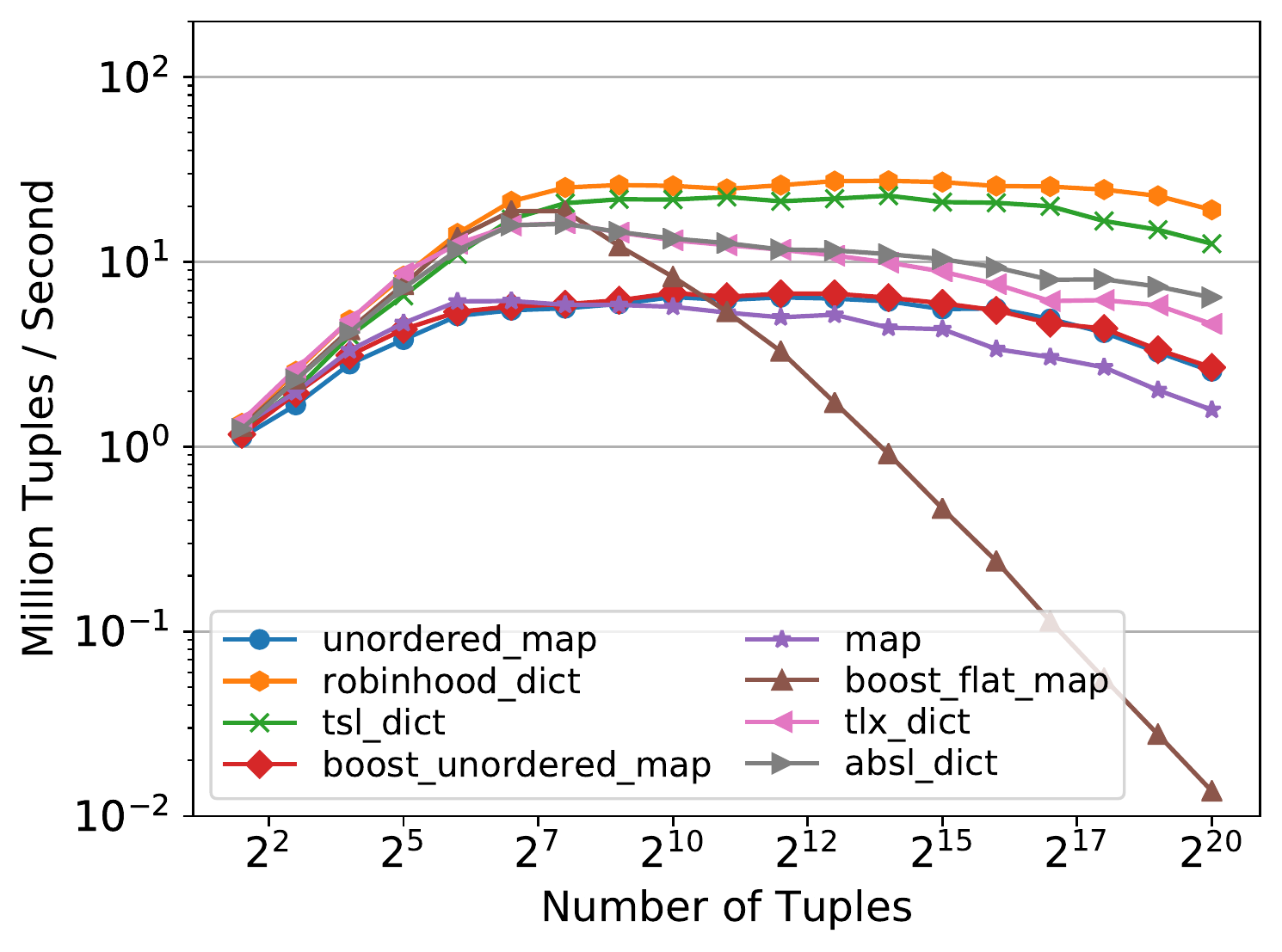}
\end{subfigure}
&
\begin{subfigure}{0.42\columnwidth}
\includegraphics[width=\columnwidth]{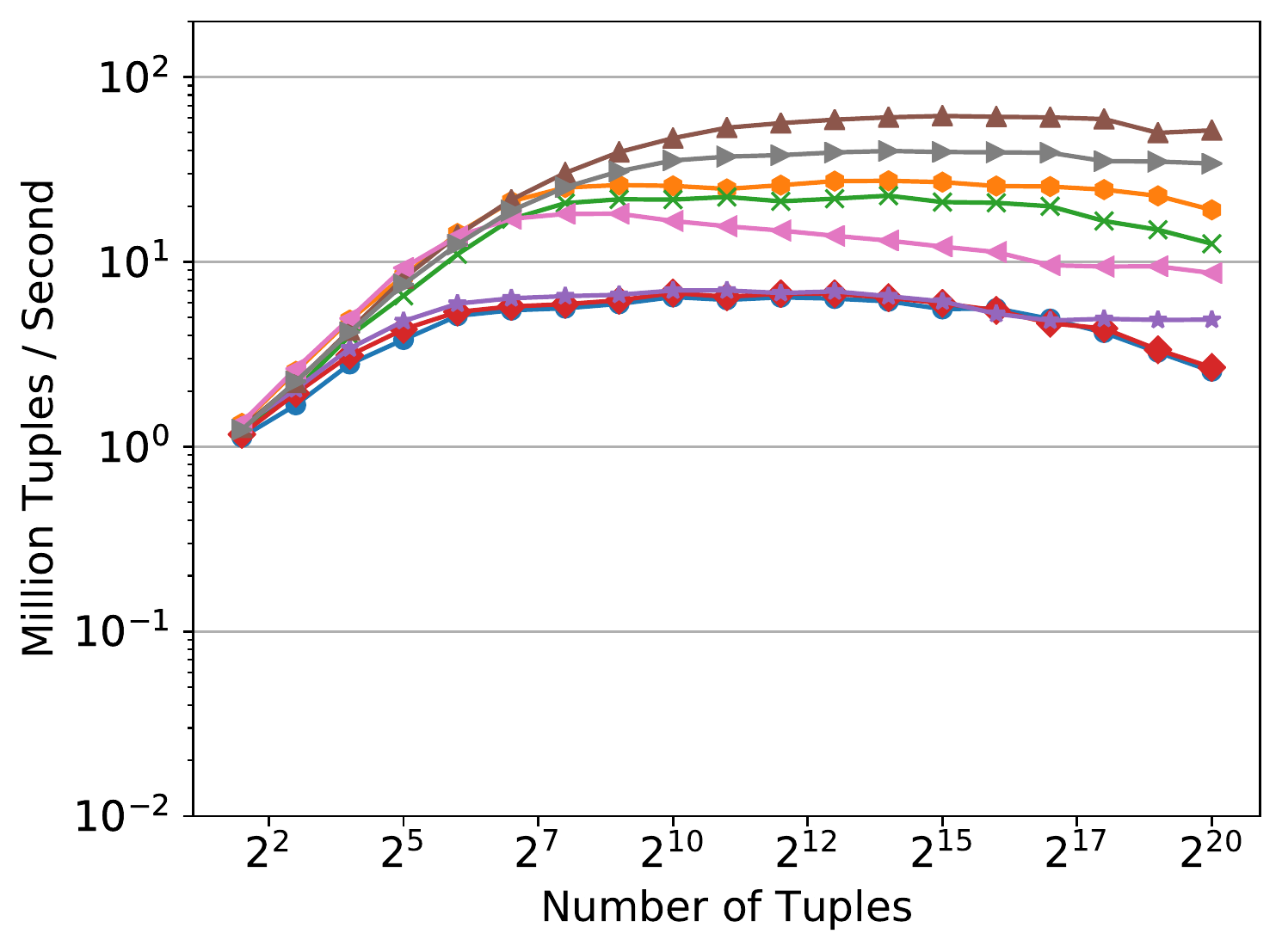}
\end{subfigure}
\\ \hline
\rotatebox{90}{\textbf{Machine 2}} &
\begin{subfigure}{0.42\columnwidth}
\includegraphics[width=\columnwidth]{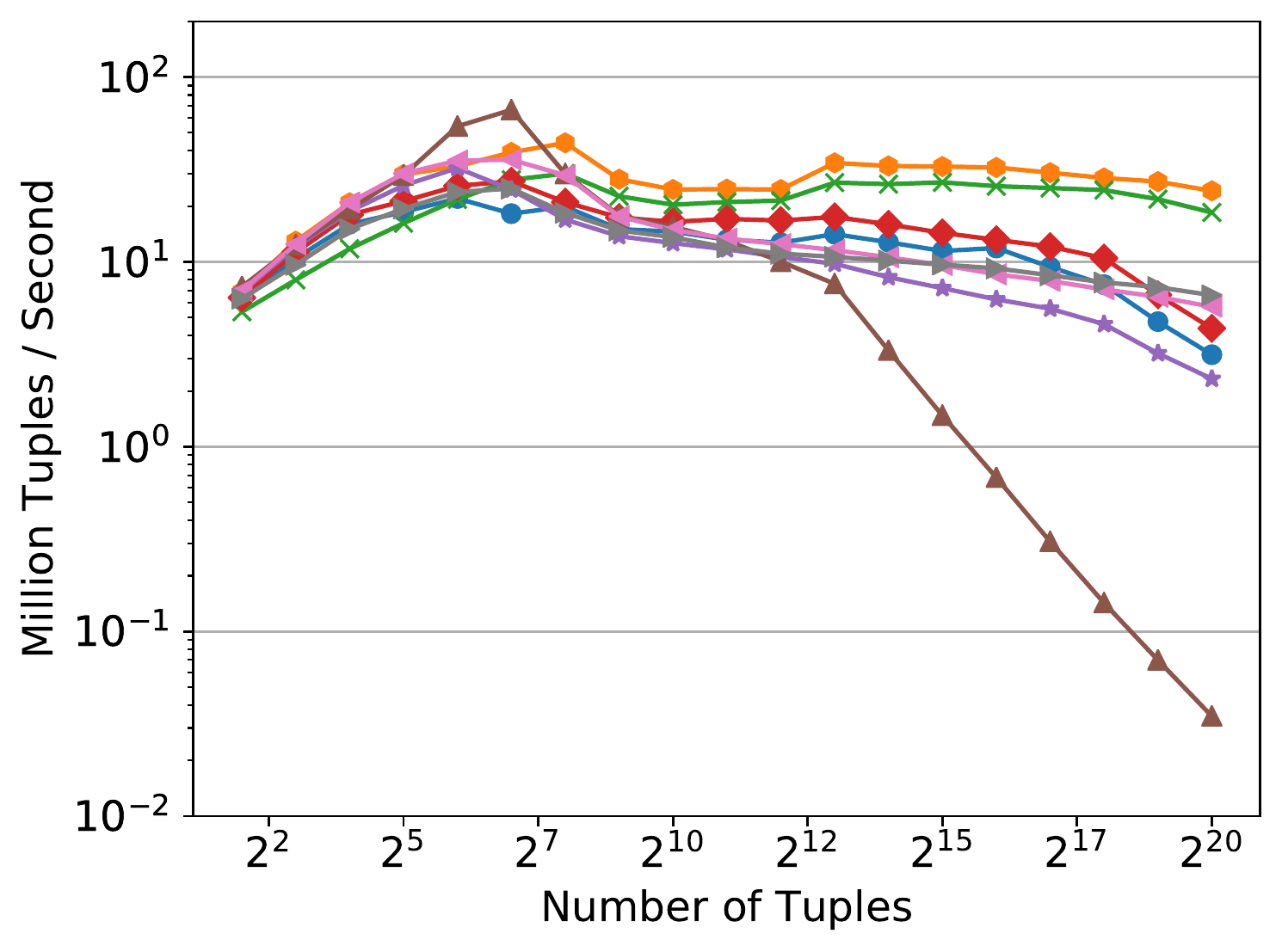}
\end{subfigure}
&
\begin{subfigure}{0.42\columnwidth}
\includegraphics[width=\columnwidth]{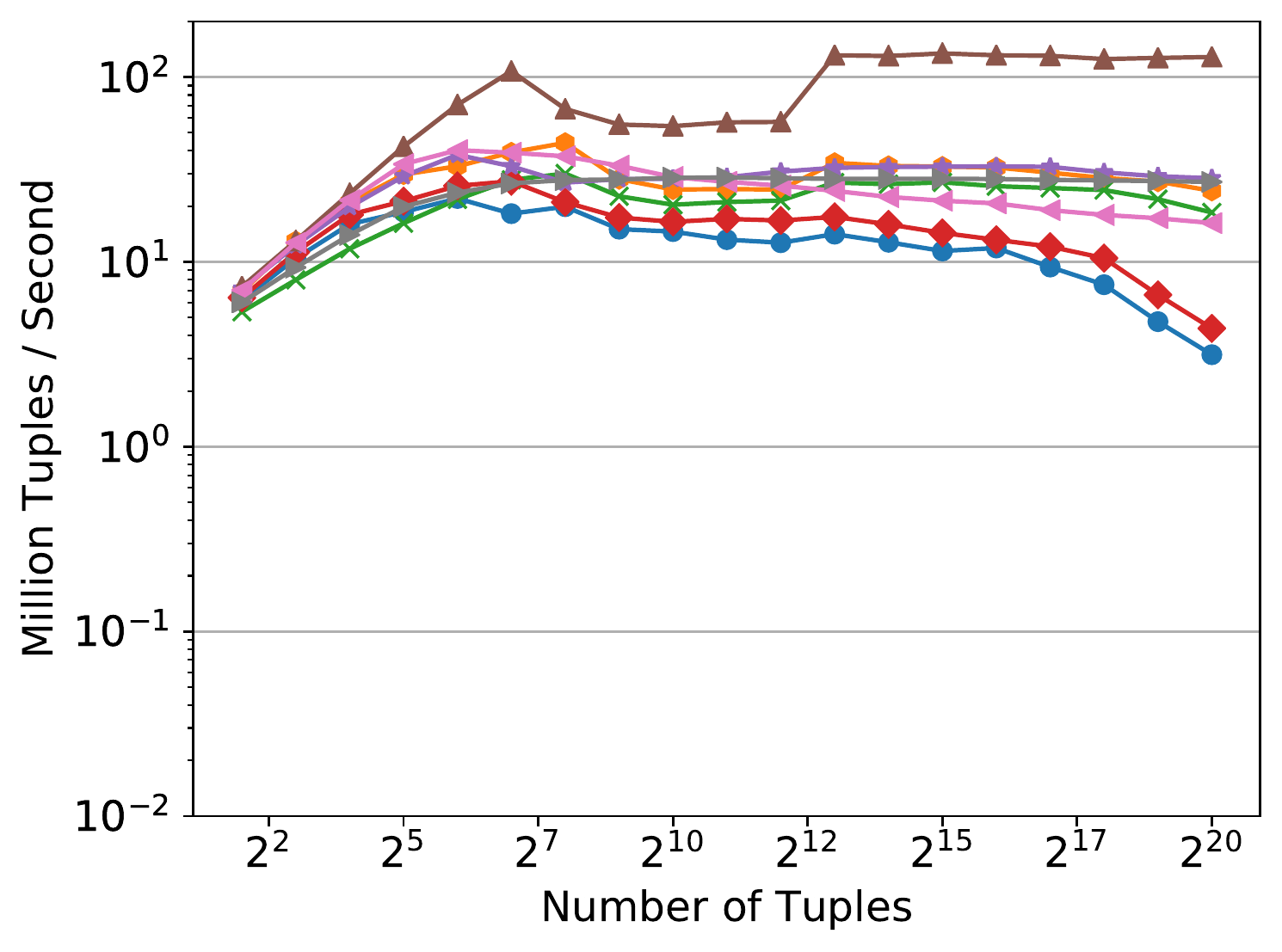}
\end{subfigure}
\\
\end{tabular}
\caption{Micro benchmark results for insert in different dictionary data structures. The key/value pairs to be inserted are either ordered based on their keys, or unordered.}
\label{fig:exp:micro_insert}
\end{figure}

Figure~\ref{fig:exp:micro_insert} shows the results for the case of insertion. 
For the case of unordered keys, we observe the superiority of hash-based dictionaries over sort-based ones. 
However, for ordered keys, the sort-based dictionaries perform better than most hash-based ones. 
An interesting case is \dsdict{\boostfmap{}}, which behaves poorly for the case of unordered keys, due to the linear insertion needed for bigger keys.
Nevertheless, this data structure outperforms others for ordered keys.

\begin{figure*}[t]
\begin{tabular}{m{1em}|c|c|c|c}
& \textit{\small Unordered / Tuples=1M} & \textit{\small Ordered / Tuples=1M} & \textit{\small Unordered / Size=16K} & \textit{\small Ordered / Size=16K} \\ \hline
\rotatebox{90}{\textbf{Machine 1}} &
\begin{subfigure}{0.45\columnwidth}
\includegraphics[width=\columnwidth]{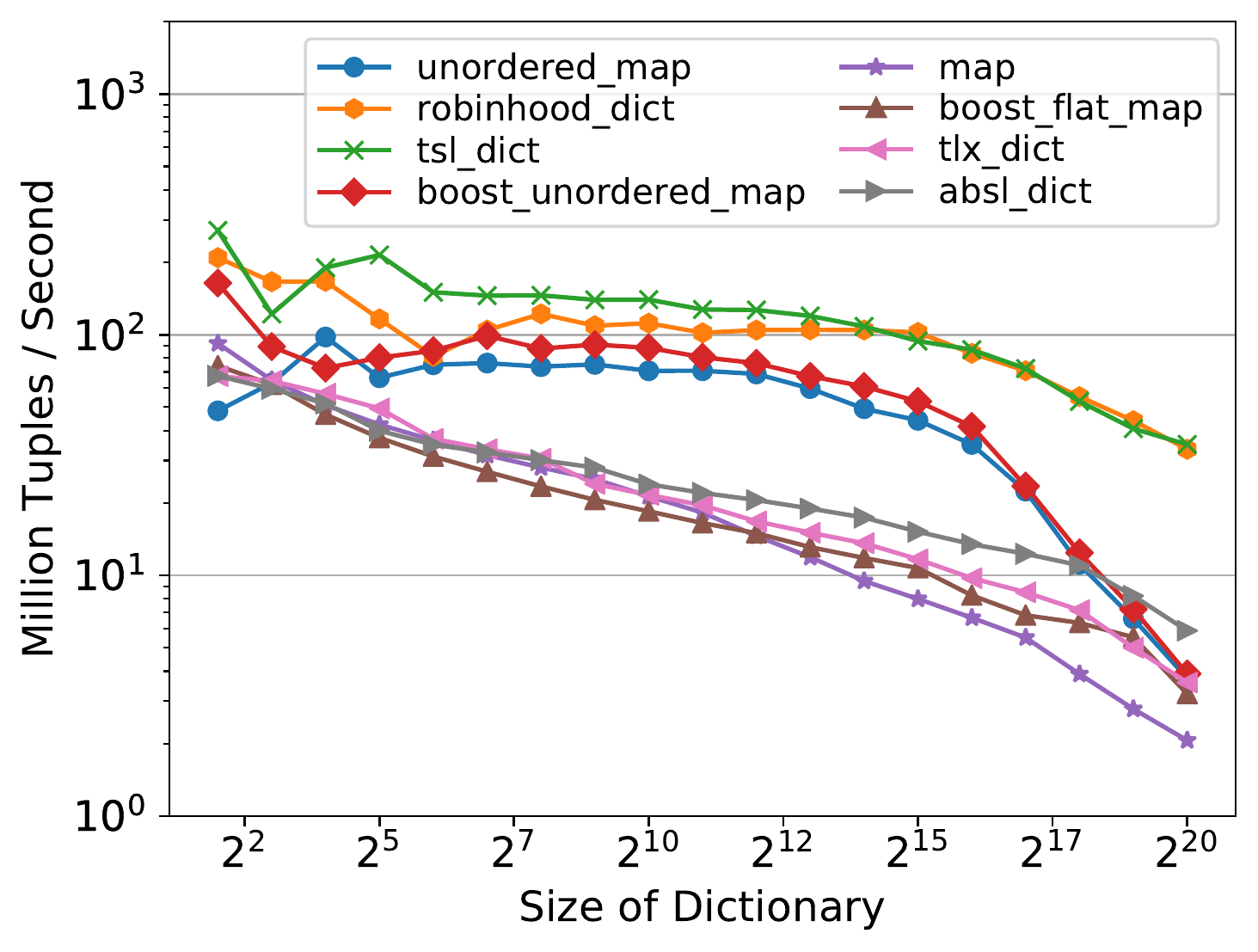}
\end{subfigure}
&
\begin{subfigure}{0.45\columnwidth}
\includegraphics[width=\columnwidth]{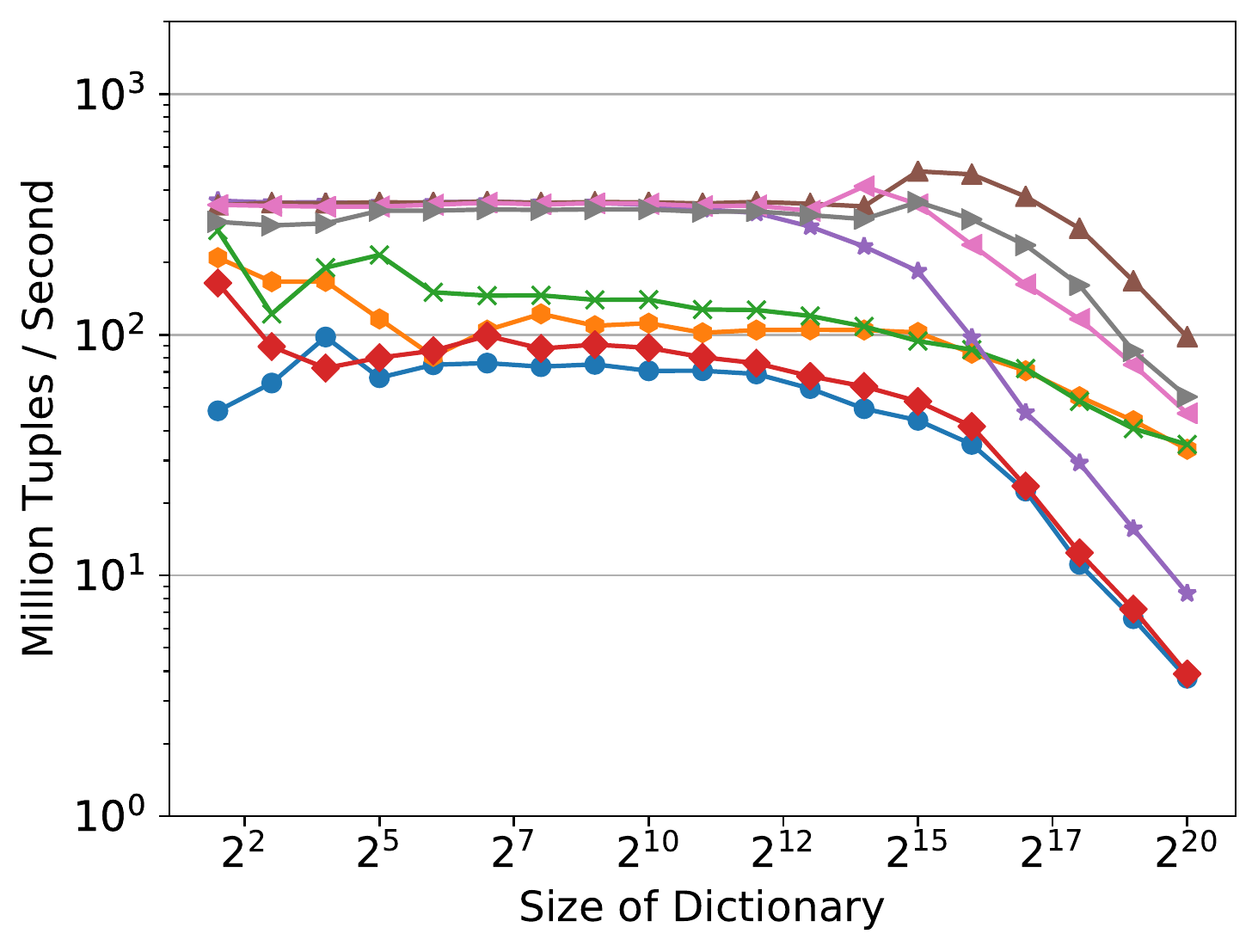}
\end{subfigure}
&
\begin{subfigure}{0.45\columnwidth}
\includegraphics[width=\columnwidth]{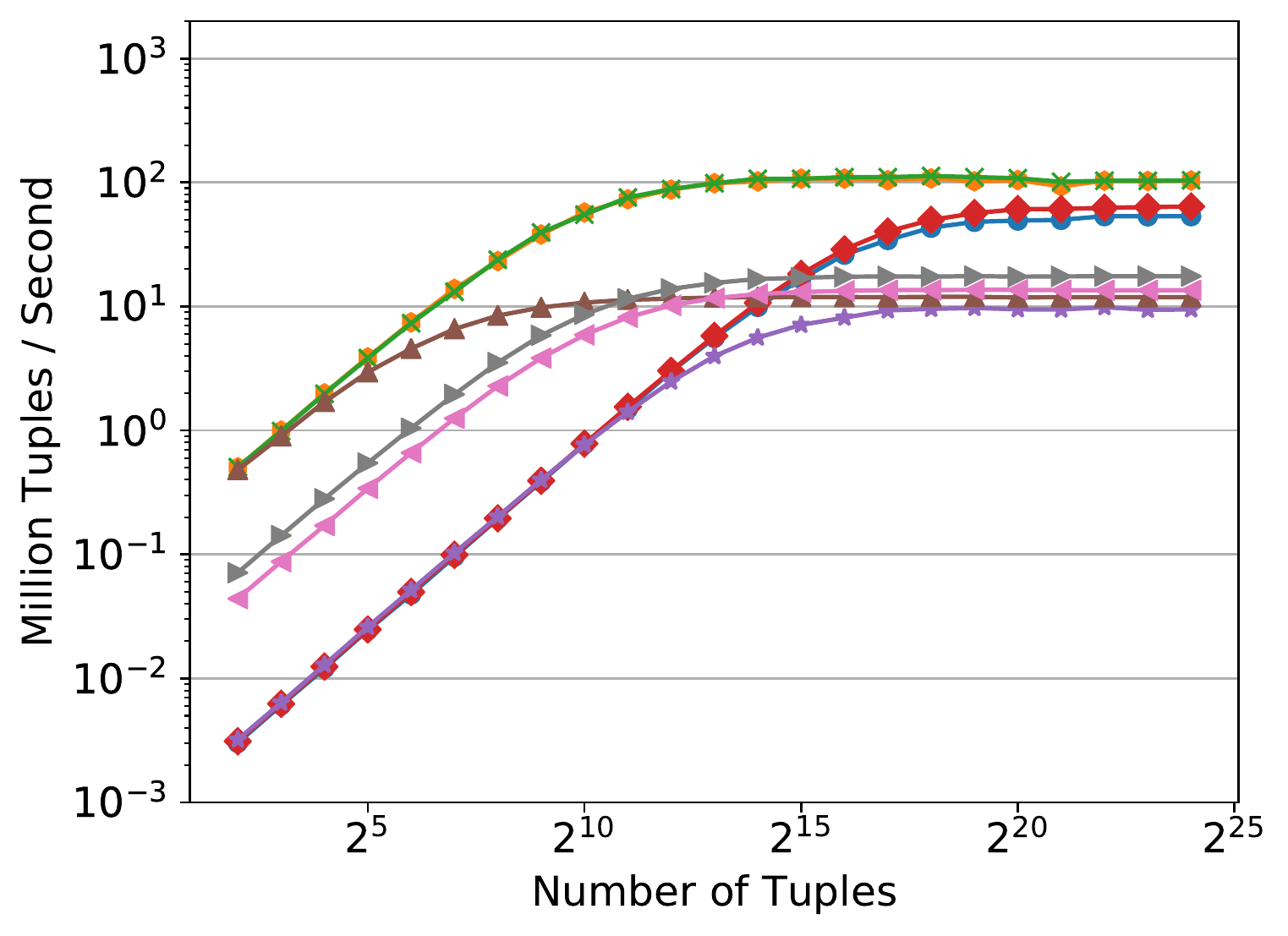}
\end{subfigure}
&
\begin{subfigure}{0.45\columnwidth}
\includegraphics[width=\columnwidth]{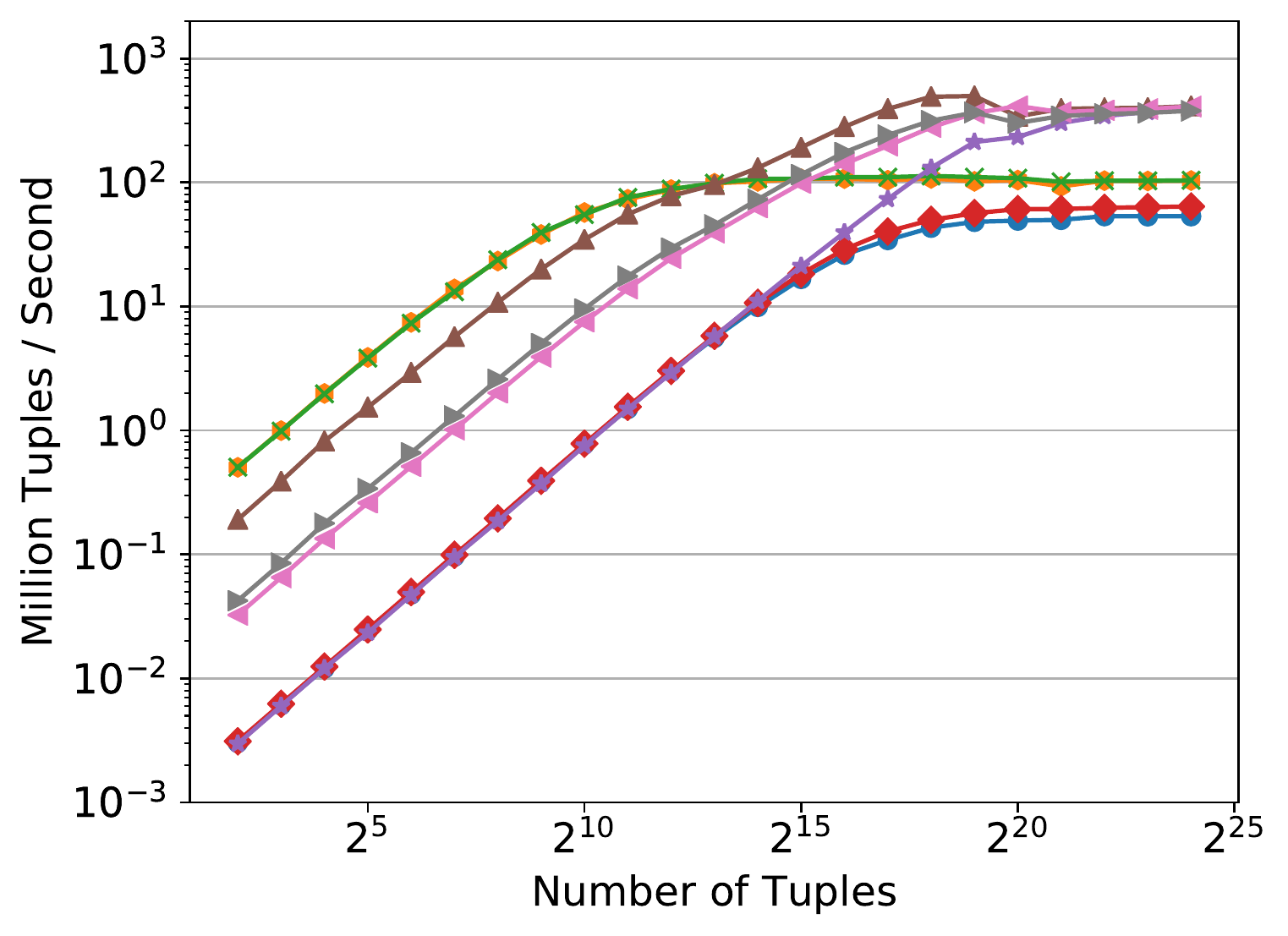}
\end{subfigure}
\\ \hline
\rotatebox{90}{\textbf{Machine 2}} &
\begin{subfigure}{0.45\columnwidth}
\includegraphics[width=\columnwidth]{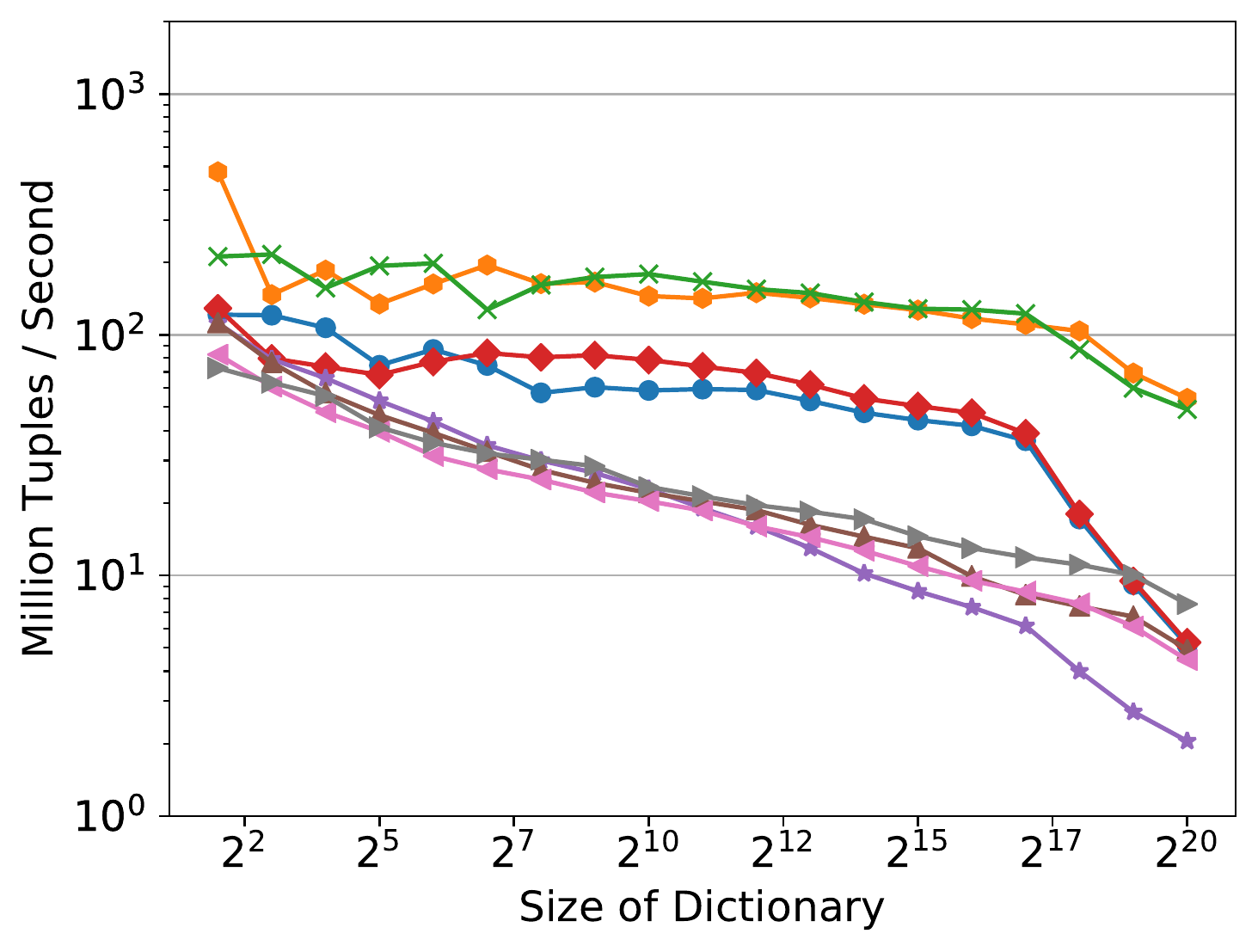}
\end{subfigure}
&
\begin{subfigure}{0.45\columnwidth}
\includegraphics[width=\columnwidth]{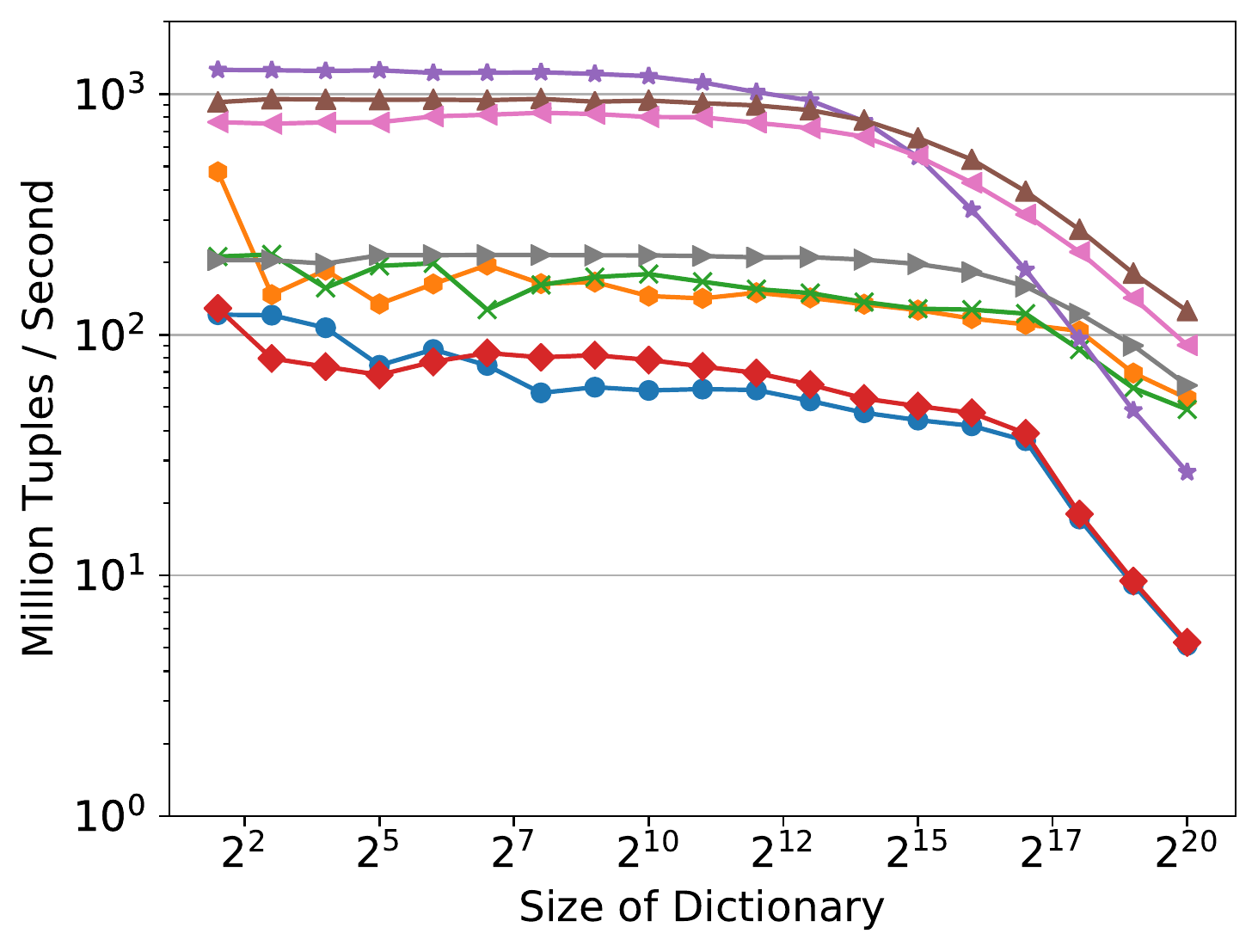}
\end{subfigure}
&
\begin{subfigure}{0.45\columnwidth}
\includegraphics[width=\columnwidth]{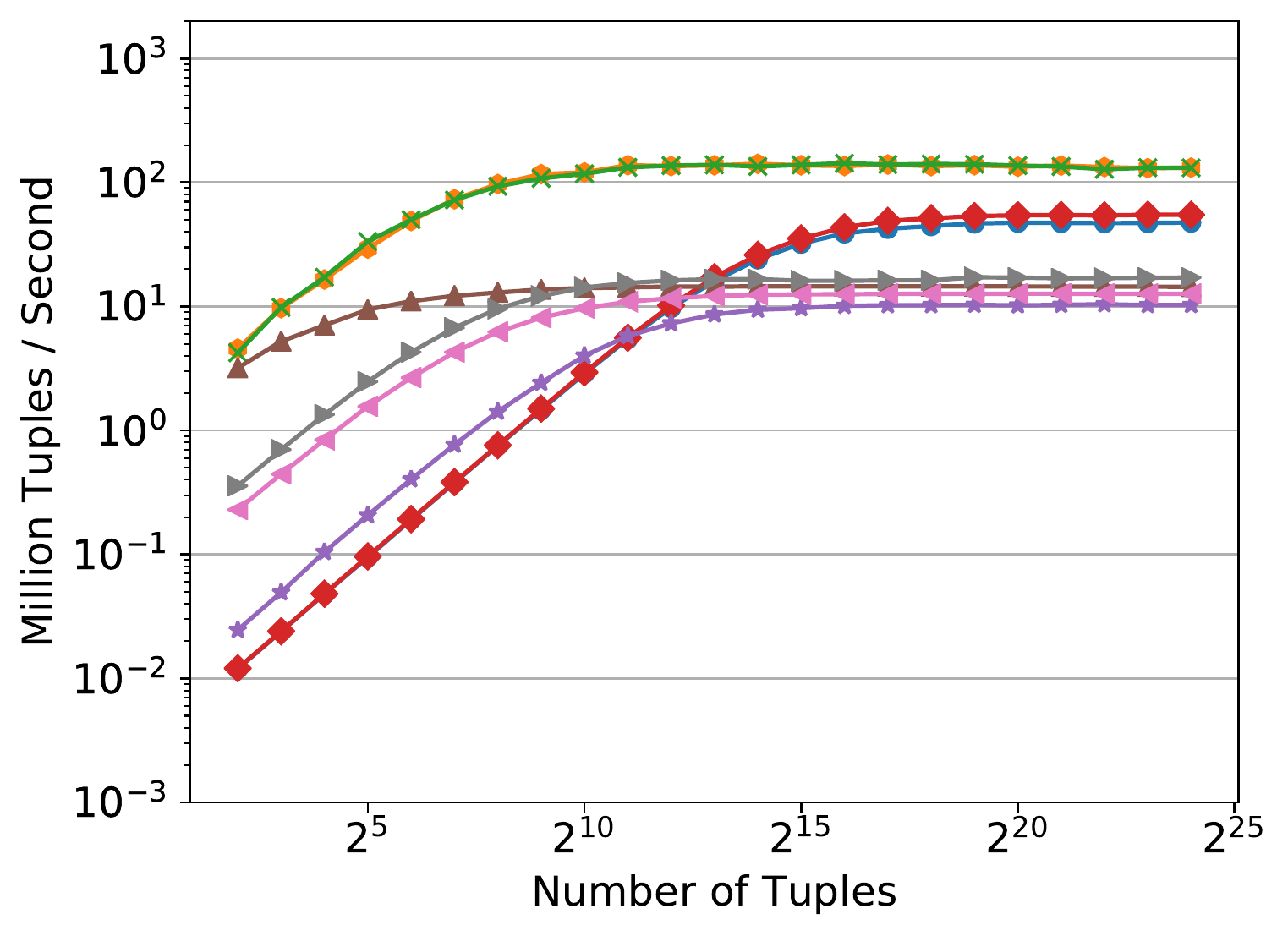}
\end{subfigure}
&
\begin{subfigure}{0.45\columnwidth}
\includegraphics[width=\columnwidth]{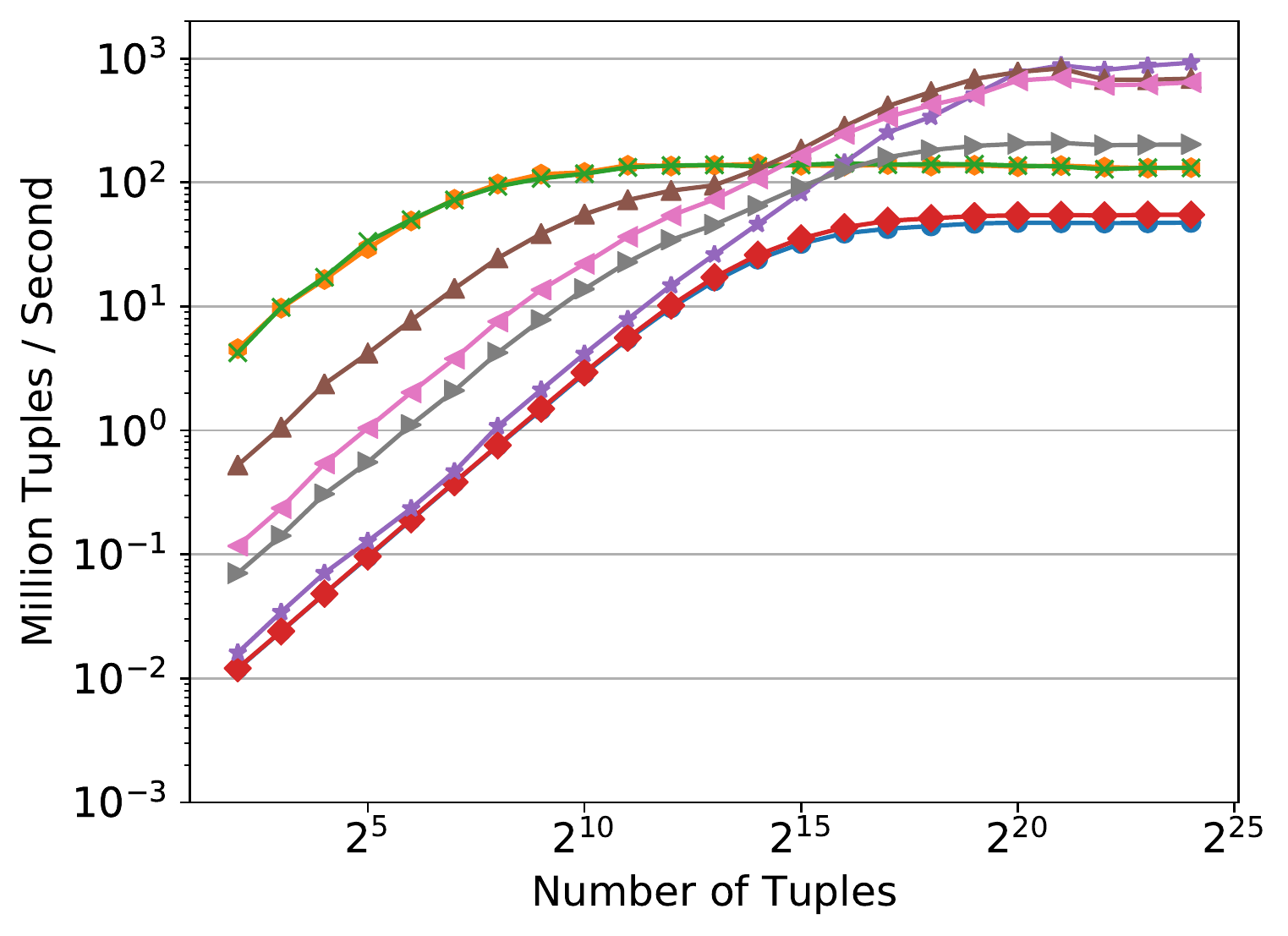}
\end{subfigure}
\\
\end{tabular}
\caption{Micro benchmark results for successful lookups in different dictionary data structures.}
\label{fig:exp:micro_lookup}
\vspace{-0.3cm}
\end{figure*}

As previously mentioned in Section~\ref{sec:ds_synthesis}, the performance of lookup operation is different
in the case of a successful lookup and unsuccessful one.
Figure~\ref{fig:exp:micro_lookup} shows the performance of successful lookup operations.
Apart from unordered and ordered data, to better demonstrate the behavior of different dictionaries, we vary both the dictionary size and the number of lookup invocations.
Similarly to the insert operation, for unordered data one observes the superiority of the hash-based dictionaries.
For ordered data, the sort-based dictionaries outperform the hash-based ones, and the performance gap widens as the number of lookup invocations increases. 
One exception is \dsdict{\stlmap{}}, for which the inefficient tree traversal leads to worse performance.
Furthermore, for a fixed dictionary size, decreasing the number of lookup invocations 
may make the amortized hinted lookup cost more than the non-hinted lookup.
We later observe in Section~\ref{sec:exp:indbml} the cases where using non-hinted lookups is preferred over hinted ones.

\begin{figure}[h]
\begin{tabular}{m{0.5em}|c|c}
& \textit{\small Unordered / Tuples=1M} & \textit{\small Unordered / Size=16K} \\ \hline
\rotatebox{90}{\textbf{Machine 1}} &
\begin{subfigure}{0.42\columnwidth}
\includegraphics[width=\columnwidth]{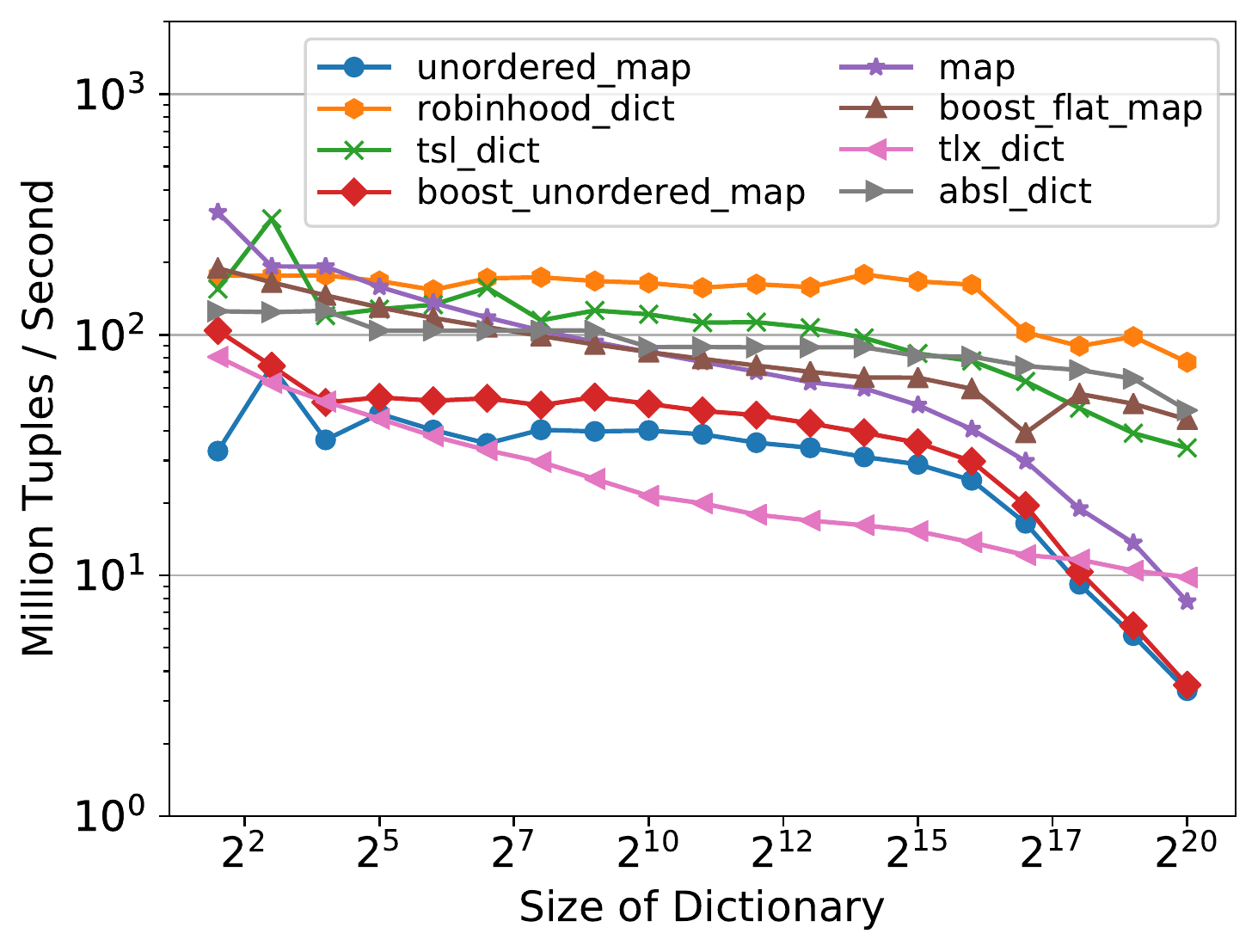}
\end{subfigure}
&
\begin{subfigure}{0.42\columnwidth}
\includegraphics[width=\columnwidth]{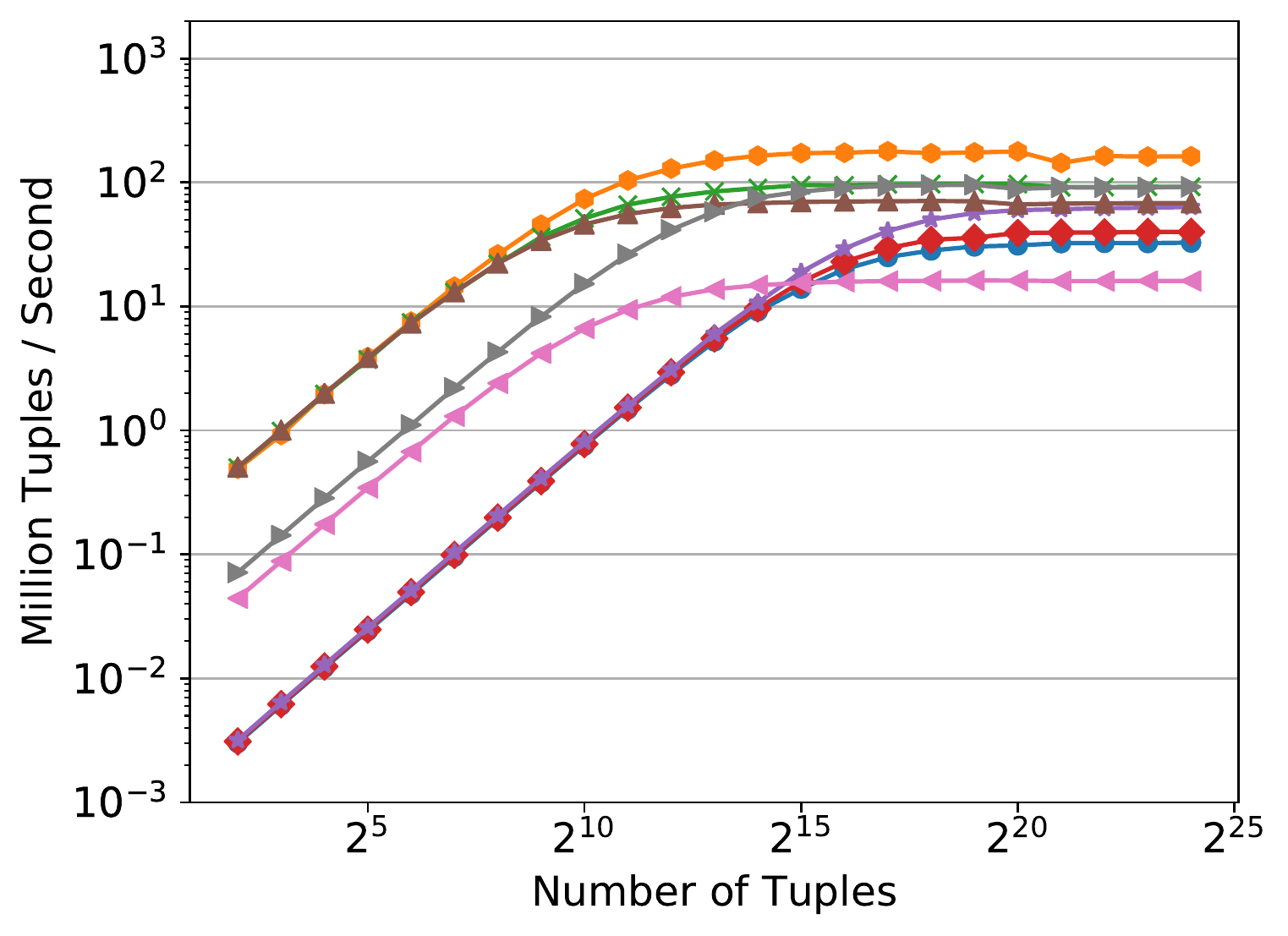}
\end{subfigure}
\\ \hline
\rotatebox{90}{\textbf{Machine 2}} &
\begin{subfigure}{0.42\columnwidth}
\includegraphics[width=\columnwidth]{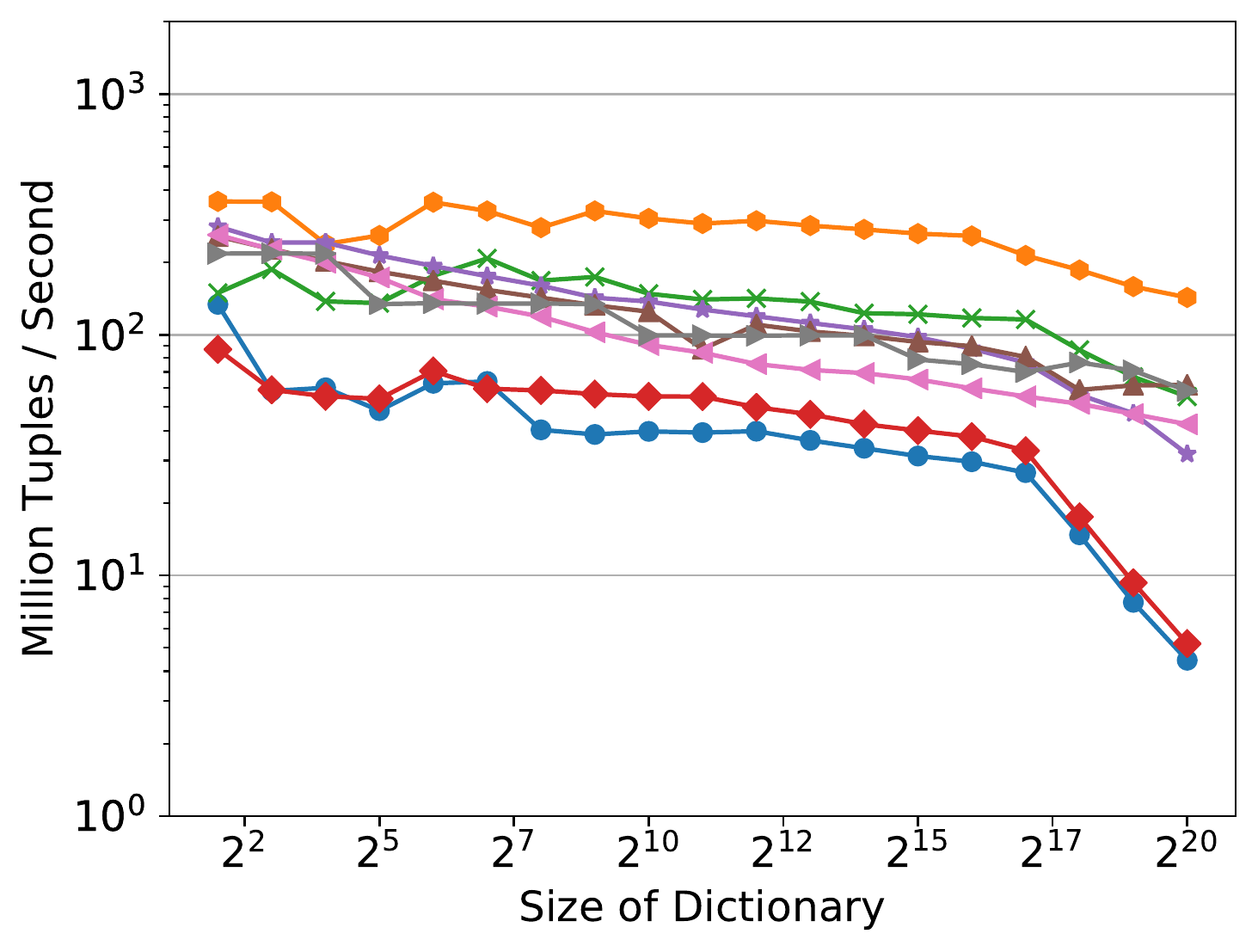}
\end{subfigure}
&
\begin{subfigure}{0.42\columnwidth}
\includegraphics[width=\columnwidth]{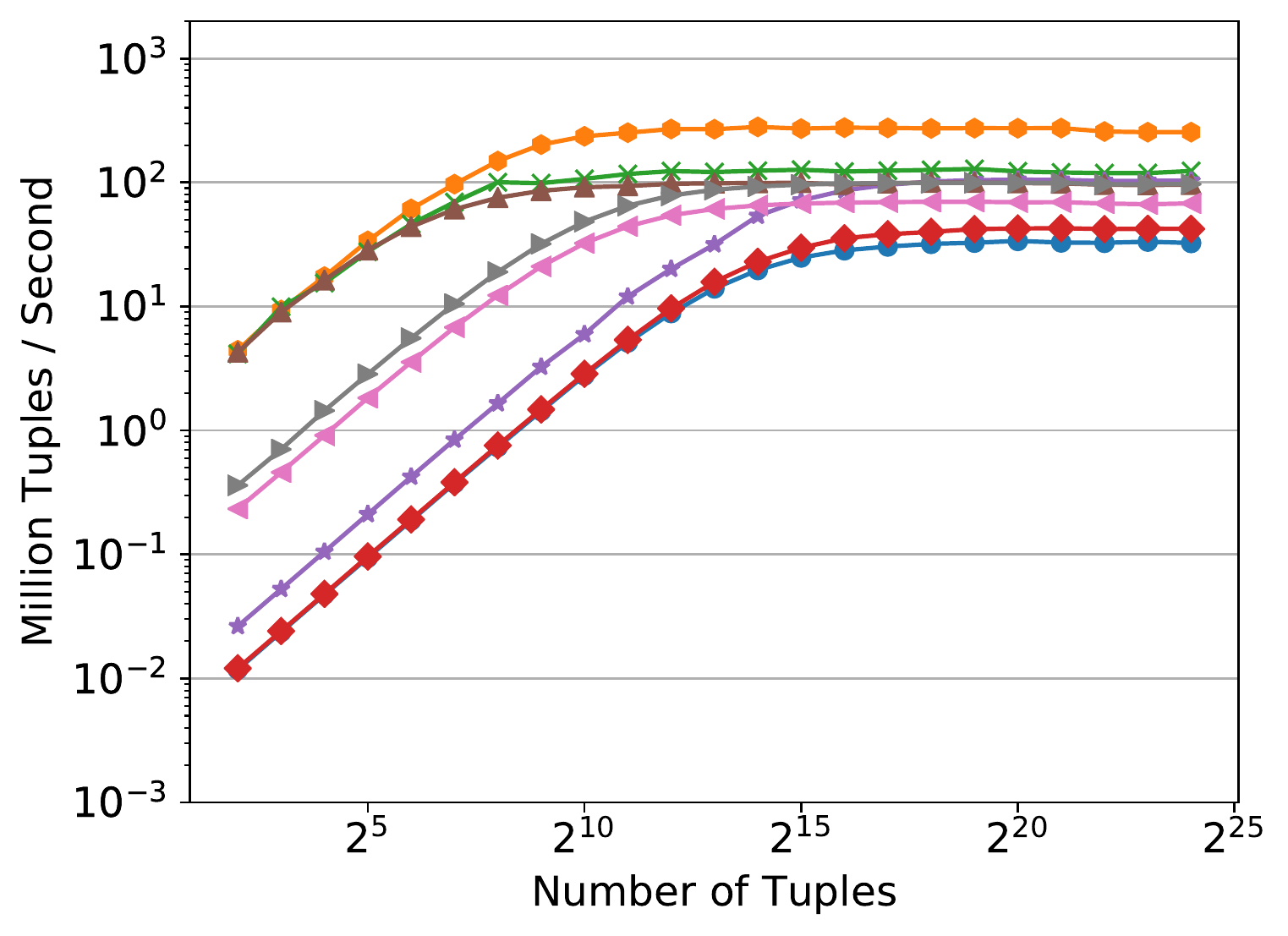}
\end{subfigure}
\\
\end{tabular}
\vspace{0.1cm}
\caption{Micro benchmark results for unsuccessful lookups in different dictionary data structures.}
\label{fig:exp:micro_lookupf}
\vspace{-0.5cm}
\end{figure}

Figure~\ref{fig:exp:micro_lookupf} shows the results for unsuccessful lookups.
Although for insertion and successful lookup operations both \dsdict{\tsldict{}} and \dsdict{\rbdict{}} behave similarly, for unsuccessful lookups the latter dictionary clearly shows a better behavior. 
For ordered keys, the performance are similar to successful lookups, thus, omitted from the experiments.

Note that all figures use a logarithmic scale for both axes. For both insertion and
lookup we observe up to three orders of magnitude performance difference among the
different dictionary implementations. This is interesting because all the implementations
are stable and used massively in software artifacts written using C++.

Finally, we observe that for all operations, the micro-architecture of the underlying 
hardware has an impact on the relative performance of different 
dictionary implementations. Here we give three important examples. 
First, \dsdict{\boostfmap} 
performs better than hash-based dictionaries on insertions for dictionary size of smaller than 100 elements 
in machine 2, but not in machine 1. 
Second, the successful lookup operator of \dsdict{\stlmap} for ordered data is faster than
all other dictionaries for dictionary size of smaller than 8000 elements in machine 2, but not in machine 1.
Third, the unsuccessful lookup operator of \dsdict{\rbdict} is consistently better than other dictionaries for every dictionary size in machine 1, but not in machine 2.

\section{Regression Models}
\label{sec:regr}
In this section, we investigate the space of possible regression models trained using scikit-learn. All of the models are trained based on all three methods explained in section~\ref{sec:exp:cost}:
\begin{enumerate*}
\item \regmodel{Linear}: linear regression.
\item \regmodel{Polynomial}: degree-2 polynomial regression.
\item \regmodel{SVM}: linear support-vector machine.
\item \regmodel{KNN}: K-nearest neighbor (K = 4).
\item \regmodel{Decision Tree}: regression decision tree of depth 5.
\item \regmodel{AdaBoost}: AdaBoost~\cite{freund1997decision} with 200 estimators.
\item \regmodel{Gradient Boost}: gradient boosting with 200 estimators.
\item \regmodel{Random Forest}: random forest with 200 estimators.
\end{enumerate*}

\begin{figure*}[t]
\includegraphics[width=2.1\columnwidth]{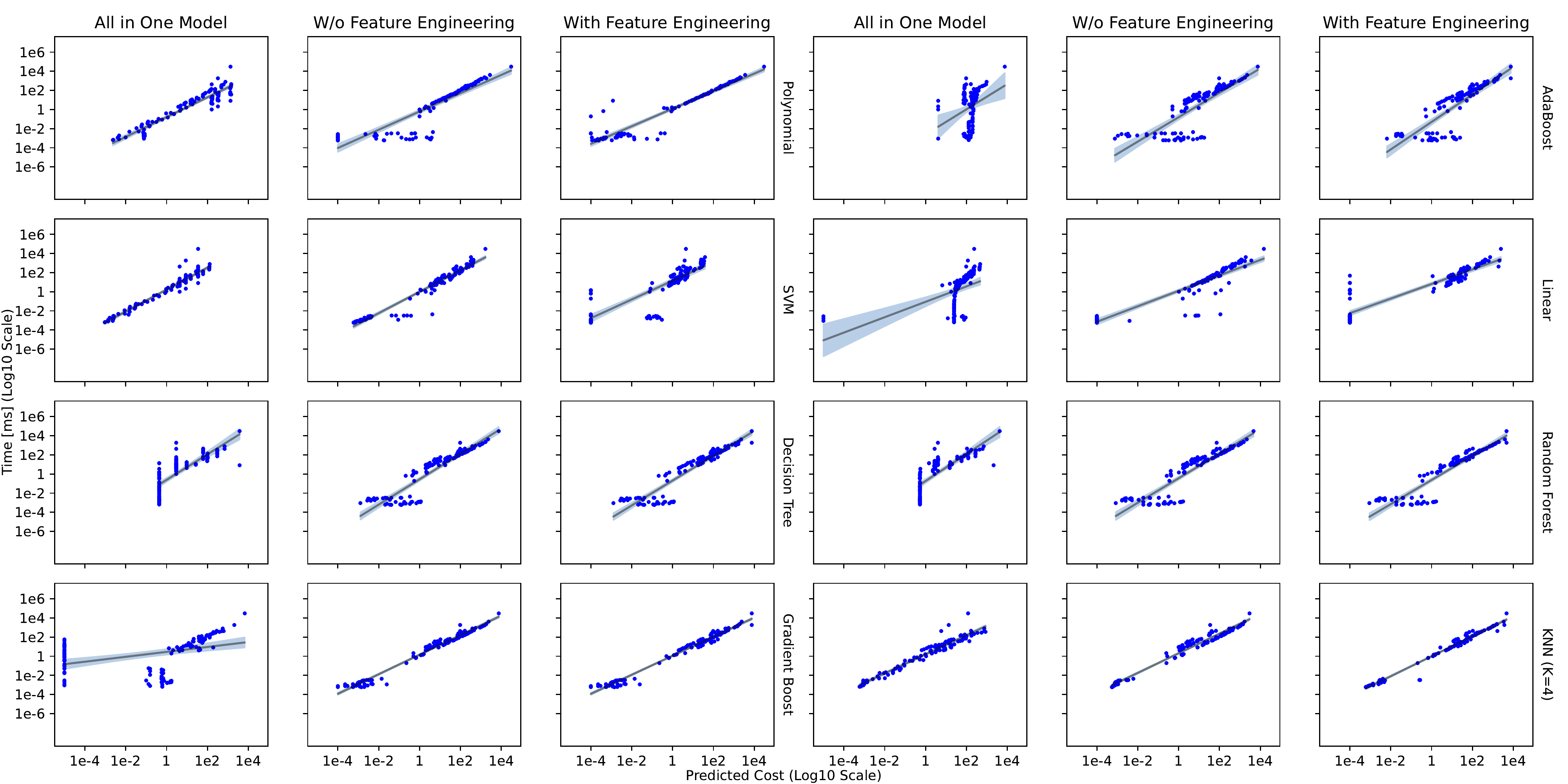}
\caption{Comparison of the prediction of 8 different regression models trained under various methods with operation running times.}
\label{fig:regrmodels}
\end{figure*}

\noindent
Figure~\ref{fig:regrmodels} shows the relation between the predicted and actual running times for operations such as lookup and insert. Utilizing 32 models (with and without feature engineering) outperforms the All in One Model training method. Splitting the cost prediction task among several models and providing them simpler integral feature values (not one-hot encoded) makes the estimation easier. Therefore, such performance is expected.

The \regmodel{KNN} model training with logarithmic features performs the best among all these models.
It uses as feature the logarithm of dictionary size which captures accurately the relationship between dictionary size and actual operation cost.
For the same reason, \regmodel{Polynomial} model which trained with similar method behaves better than \regmodel{Linear} and \regmodel{Polynomial} models W/o feature engineering.
This may be unsurprising, as the expected cost is logarithmic in dictionary size for sort-based dictionaries. 
Yet the constant factors in the complexity are now captured more accurately by the model.
Among the tree-based models, \regmodel{Gradient Boost} outperforms others.
and trained them using scikit-learn~\cite{pedregosa2011scikit}